\documentclass[12pt,draftclsnofoot, onecolumn]{IEEEtran}
\usepackage{amssymb}
\usepackage{amsthm}
\usepackage{graphicx}
\usepackage{array,color}
\usepackage{amsmath}

\usepackage{graphicx}
\usepackage{tabularx}
\usepackage{epsfig,epsf,color,balance,cite}
\usepackage{algorithmic}
\usepackage{algorithm}
\usepackage{bm}
\usepackage{caption}
\usepackage{lipsum}
\usepackage{cuted}

\usepackage{textcomp}
\usepackage{color}
\usepackage{multirow}
\usepackage{cite}
\usepackage{enumerate}
\usepackage{cases}
\usepackage{color}
\usepackage{url}
\usepackage{epstopdf}
\usepackage{textcomp}
\usepackage{subcaption}
\usepackage{hyperref}

\newcounter{MYtempeqncnt}

\begin{document}

\title{RIS-Aided Near-Field Localization and Channel Estimation for the Sub-Terahertz System}
\author{\IEEEauthorblockN{Yijin Pan, Cunhua Pan, Shi Jin and Jiangzhou Wang,~\IEEEmembership{Fellow,~IEEE}}
\thanks{Y. Pan, C. Pan and S. Jin are with the National Mobile Communications Research Laboratory, Southeast University, Nanjing 211111, China.  (Corresponding author: Cunhua Pan) }
\thanks{J. Wang is with the School of Engineering, University of Kent, Canterbury, CT2 7NT, UK.}}
\maketitle

\begin{abstract}
	The low hardware cost makes ultra-large (XL) reconfigurable intelligent surface (RIS) an attractive solution for the performance enhancement of localization and communication systems, but it results in near-field propagation channels, especially for the high-frequency communication systems, e.g. mmWave and sub-terahertz (subTHz) systems.
	This makes localization and channel estimation in the near field of the RIS-aided subTHz system more challenging.
	In this paper, we consider the spherical wavefront propagation in the near field of the subTHz system with the assistance of a RIS.
	A near-field channel estimation and localization (NF-JCEL) algorithm is proposed based on the derived second-order Fresnel approximation of the near-field channel model.
	To be specific, we first decouple the user equipment (UE) distances and angles of arrival (AoAs)  through a down-sampled Toeplitz covariance matrix, so that the vertical and azimuth AoAs in the array steering vectors can be estimated separately with low complexities.
	Then, the UE distance can be estimated by the simple one-dimensional search, and the channel attenuation coefficients are obtained through the orthogonal matching pursuit (OMP) method.
	Simulation results validate the superiority of the proposed NF-JCEL algorithm to the conventional far-field algorithm, and show that higher resolution accuracy can be obtained by the proposed algorithm.
\end{abstract}


\section{Introduction}

The sixth-generation (6G) mobile networks are expected to cater to various Internet of Things (IoT) applications such as healthcare, smart homes and intelligent manufacture\cite{you2021towards}.
These applications and services entail decision making on the basis of high-resolution location data from the involved collections of sensors, machines and other items.
For 6G services, sub teraherze (subTHz) spectrum, i.e. from 90 GHz to 300 GHz,  has primarily been a driving force for high-resolution sensing with centimeter-level positioning accuracy\cite{kanhere2021position,9558848}, resulting from its high frequency, sufficiently large bandwidth and increased spectral efficiency\cite{ning2021terahertz,pan2022sum}.

However, radio transmission at high frequency, for instance subTHz spectrum, suffer from high path loss, bad propagation conditions and is easy to be blocked.
To cope with this issue, the reconfigurable intelligent surfaces (RISs) has been regarded as a promising solution to overcome the blockage issue of the subTHz localization system.
By carefully tuning the phase shifts of the reflecting elements, RIS can help constructively accumulate increased power at the target receiver, and thus shows a great potential to construct the intelligent radio environment.
Thus, it was leveraged to help enhance performance of the simultaneous wireless information and power transfer (SWIPT) system \cite{9110849,pan2022self} and multicell network \cite{9090356}.
Specially, in localization system, it is shown that the RIS can help make the localization more accurate\cite{wu2022two,9847080}.
It was shown in \cite{Wang.202131} that proper RIS phase shifts configuration can also help improve the Cram$\rm{\acute{e}}$r-Rao lower bound (CRLB) of the position/orientation estimation.

Conventionally, the UE localization problems aim to utilize the received signal strength (RSS), the time-difference-of-arrival (TDoA)\cite{Chan.1994}, and the angle of arrival (AoA) of multiple sensors to distinguish the multiple signal directions from different emitters.
The main idea is that all the geometric information for localization is included in these channel state information (CSI) measurements.
Thus, the RIS-assisted localization accuracy is highly dependent on the necessity to acquire side channel measurements \cite{Wymeersch.20191220}.
However, the channels between the RIS and UEs, as well as the channel between the access point (AP) and the RIS, are cascaded\cite{Zhang.2022125}.
In addition, the RIS has little signal processing capabilities, making channel estimation of RIS-assisted channels a challenging task\cite{Pan.20211211}.
Many researches successfully managed to obtain the cascaded channel gains, such as \cite{Zhou.2021628}, which have fully exploited the correlations between UEs and different paths to reduce the pilot overhead.
In the RIS-assisted mmWave localization system, beam training was investigated in \cite{Wang.2021} to estimate AoAs and angles of departure (AoDs) of the line of sight (LoS) paths.
In \cite{9456027}, the RSS was utilized in the RIS aided localization system, and the RIS phase shifts were optimized to minimize the weighted probabilities of false localization.
The impact of phase shifts and the number of reflecting elements on the CSI parameter estimation, positioning error bound (PEB) and orientation error bound (OEB) was studied in \cite{9129075}.

However, the above researches were mainly based on the assumption of far field channel model, where the planar wavefront was adopted.
Thanks to the low hardware cost and reduced power consumption, a large number of passive and reflecting elements can be integrated into an RIS panel, so that its size can reach several meters.
In some envisioned industrial IoT application scenarios, the RIS panels can be installed to cover the entire roof and walls of a building.
However, with the extra-large RIS panel, the spherical wavefront feature should be considered, which is the so-called the near-field effect.
According to \cite{selvan2017fraunhofer}, to maintain a maximum phase difference of $\pi/8$ rad, the observation distance from the UE/AP to the RIS must be no less than the Fraunhofer distance $\frac{2L^{2}}{\lambda}$, where $L$ and $\lambda$ denote the maximal aperture of the RIS panel (i.e. the diagonal length of a rectangular panel) and radio wavelength, respectively.
For instance, for an RIS panel with aperture $50$ cm, the Fraunhofer distance is $83.33$ m for the electrical magnetism (EM) waves at $50$ GHz ($\lambda = 6$ mm), and it is increased to $166.67$ m for the radio frequency increasing to $100$ GHz ($\lambda = 3$ mm) in THz band.
Therefore, in this case, the UEs are very likely in the near field of the RIS panel and the general spherical wavefront should be considered.  
With the spherical wavefront radio, the emitted EM wave will arrive at each reflecting element in the RIS array with different AoAs.
This indicates that all array elements share a common AoA on the same path in the far-field model is no longer valid \cite{Chen.2002}.
The near-field effect makes the channel modeling and CSI estimation for localization more challenging, and the near-field communications will profoundly degrade the localization performance\cite{Friedlander.2019b}, as also to be proved in this paper.

The study on the RIS behavior in the near-field is just in its infancy.
The RIS reflected power behavior was analyzed and measured in \cite{Tang.2021121} under near/far-field conditions.
The near-field channel modeling for active antenna arrays and RIS was investigated in \cite{Bjornson.20211013}.
In \cite{9617121}, a generic communication model with XL-array/surface was investigated with the consideration of the variations of signal phase, amplitude and aperture across array elements.
The channel model mismatch was addressed in \cite{chen2022channel} by leveraging the MCRB (misspecified Cramer-Rao lower bound) to lower bound the localization error using a simplified mismatched model.
A near-field codebook was developed in \cite{Wei.2021921} for the XL-RIS beam training by dividing the three-dimensional (3D) space into sampled points in the x-y-z coordinate system.
The authors of \cite{9133126} demonstrated that a RIS can serve as an anomalous mirror in the antenna array's near field, by using antenna theory to calculate the electric field of a finite-size RIS.
As for the radio localization approaches with RIS, there are only a few works with the consideration of spherical wavefront in the near-field.
In \cite{AbuShaban.62021}, the Fisher information matrix (FIM) was analyzed for an uplink localization system using a RIS-based lens, and the PEB and OEB were also evaluated by exploiting the wavefront curvature. 
The CRLB for RIS-assisted localization was investigated in \cite{Elzanaty.2021} and the RIS phase shift was optimized to maximize the received signal to noise ratio (SNR).
Similar researches were also presented in \cite{Rahal.2021924}, where FIM was analyzed for the multipath-aided localization with only RIS-assisted non line of sight (NLoS) links.

Most existing works such as \cite{Bjornson.20211013,9617121,chen2022channel} considered the near-field channel modeling and system performance analysis but did not address the issue of angle estimation and positioning algorithm in the near field.
Although several RIS-assisted localization schemes have been proposed for the near-field case\cite{AbuShaban.62021,Elzanaty.2021,9921216}, they were focused on the estimation error bound analysis, such as FIM/PEB/CRLB, and there is a lack of the specific positioning algorithm.
In addition, some works such as \cite{chen2022ris,rinchi2022compressive} considered the uniform linear array (ULA) of RIS.
However, the proposed localization solutions in these works cannot be directly applied to the UPA case, as only either the azimuth or elevation AoA/AoD was considered.
These researches are not sufficient for the more general 3D localization system, and cannot be directly applied to the RIS channel with a uniform planar array (UPA).
In practice, RIS is normally deployed as a panel, and thus the angles in both azimuth or elevation directions should be jointly estimated, however, which is not a simple extension to the linear case.
According to our work, the down-sampled Toeplitz covariance matrix should be carefully re-investigated, and angles in different directions need to be further decoupled to avoid two-dimensional grid searching, and these tasks for the UPA RIS system are not trivial.
In addition, the benefits of the adjustable phase shifts of the RIS panel are not fully exploited in current work.
For instance, the RIS phase shifts were considered to improve the error bounds in \cite{AbuShaban.62021,Elzanaty.2021,9921216,chen2022ris}, and to maximize the SNR in \cite{rinchi2022compressive}.
In fact, the RIS can be employed to efficiently increase the received signal samples from the same pilot data, thereby it has the potential to effectively improve the accuracy in the process of the localization algorithm itself.
Unfortunately, how this approach can be integrated into the localization algorithm has not been addressed.
Overall, it is still unclear about how to efficiently obtain near-field UE locations and the involved CSI parameters, such as the AoA/AoDs, in the subTHz localization system with RIS.

In this paper, we investigate the localization and CSI estimation scheme in the near field of the subTHz system with the assistance of a RIS panel. 
Our contributions are summarized as follows:
\begin{itemize}
	\item  By considering the propagation of the spherical wavefront across the RIS array, the near-field UE-RIS uplink channel is modeled.
	To reduce the complexity introduced by the different spatial paths at each reflecting element,  the channel formulation based the second-order Taylor approximation is derived, which involves both the UE distance and the AoAs.
	Then, the cascaded AP-RIS-UE channel in the near field is presented\footnote{The simplified approximation of near-field channel model is presented in conference version\cite{Gc2022}, but the following localization and channel estimation algorithm is only presented in this work.}.
	
	\item  The RIS training phase shifts and pilots are carefully designed to increase the channel rank so that the channel covariance matrix can be obtained by the least square (LS) estimation. Then, a down-sampled covariance matrix is derived to decouple the UE distance and the AoAs separately.
	
	\item Based on the down-sampled Toeplitz covariance matrix, the vertical and azimuth AoAs are estimated separately.
	Then, the UE distance can be estimated by the simple one-dimensional search, and the orthogonal matching pursuit (OMP) method is leveraged to obtain the channel attenuation coefficients.
	
	\item 
	Simulation results validate the effectiveness of the proposed near field joint channel estimation and localization (NF-JCEL) algorithm. Compared with the far-field algorithm, the proposed algorithm shows superiority in terms of localization and channel estimation accuracies.
	
\end{itemize}

The remainder of this paper is organized as follows:
Section II derives the near/far-field RIS-aided channel model and formulates the channel estimation problem.
Section III develops the detailed algorithm for joint channel and localization estimation.
In Section IV, the simulation results are presented to show the performance gain and the impact of system parameters, and Section V concludes the paper.

\textit{Notation}: For a vector $\bm{x}$, $|\bm{x}|$ and $(\bm{x})^T$ denote its Euclidean norm and its transpose, respectively.
For matrix $\bm{A}$, $\bm{A}^H$  $\bm{A}^{-1}$ and $\bm{A}^{\dag}$ represent the conjugate transpose, the inverse and Moore-Penrose pseudoinverse operator, respectively.
$\mathbb{C}^{M \times N}$ denotes the set of $M \times N$ complex matrix. 
$\text{diag}(\bm{x})$ represents the diagonal matrix $\bm{X}$ obtained from vector $\bm{x}$.
$\bm{a} \otimes \bm{b} $ represents the kronecker product of $\bm{a}$ and $\bm{b}$.

\section{Channel Model}	

\begin{figure}
	\centering
	\vspace{-2em}
	\includegraphics[width=0.7\textwidth]{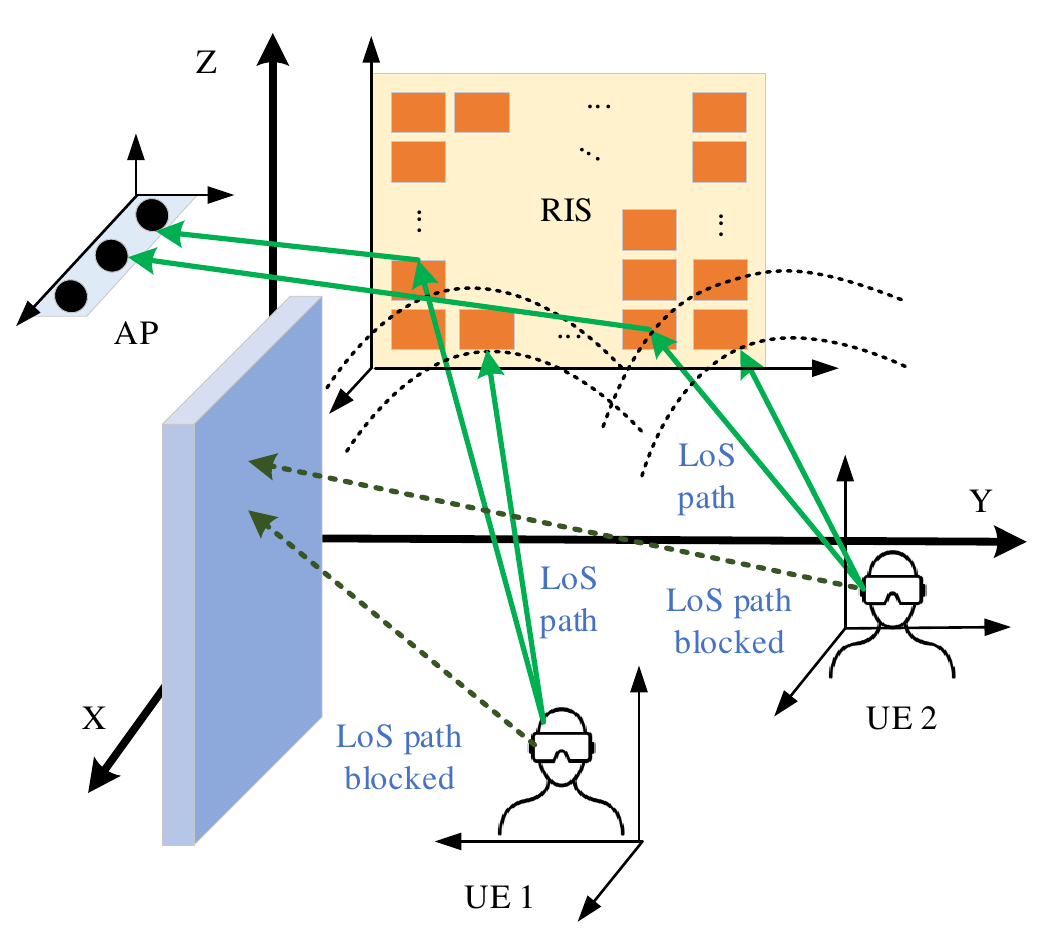}
	\vspace{-0.5 em}
	\caption{System model and communication scenario}
	\vspace{-1.5 em}
	\label{FigModel}
\end{figure}

Consider the uplink transmission of a subTHz localization system, as shown in Fig. \ref{FigModel}.
In the considered scenario, each UE is equipped with a single antenna, the AP is located in the XOZ plane, and the number of UEs is denoted as $U$. 
It is assumed that the direct links between the AP and the UEs do not exist due to the blockage or unfavorable propagation environments.
Thus, an RIS is leveraged to construct the alternative AP-RIS-UE links for localization service.

\begin{figure}
	\centering
	\includegraphics[width=0.7\textwidth]{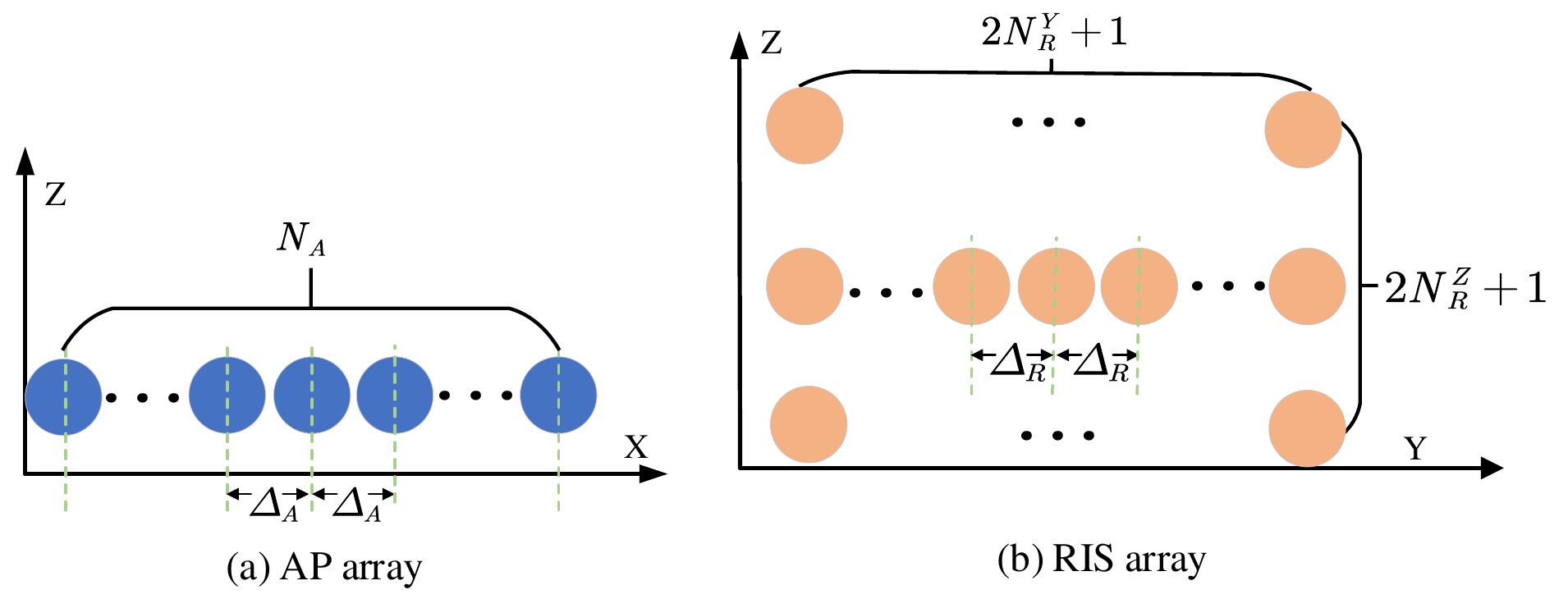}
    \vspace{-0.5 em}
	\caption{The layout of the AP antenna arrays and RIS arrays.}
	\vspace{-1.5 em}
	\label{FigAPLayout}
\end{figure}

The layouts of the AP antenna array and the RIS panel are shown in Fig. (\ref{FigAPLayout}a) and Fig. (\ref{FigAPLayout}b), respectively.
The receive antenna array of the AP is assumed to be a ULA, and the number of AP's antenna elements is denoted as $N_A$.
The distance between two antenna elements of the AP is denoted as $\Delta_A$.
As shown in Fig. (\ref{FigAPLayout}b), the RIS is installed on the wall at the YOZ-plane, and the RIS is equipped with $N_R$ reflecting elements, where the number of the reflecting elements along the Y-axis and Z-axis are denoted as $(2N^Y_{R}+1)$ and $(2N^Z_{R}+1)$, respectively.
Thus, $N_R= (2N^Z_{R}+1)\times (2N^Y_{R}+1)$.
The distance between two reflecting elements is denoted as $\Delta_R$.

Due to the short wavelength of the subTHz EM waves, second-bounce or more reflections and scattering components are severely attenuated and thus they are negligible. 
Similar to \cite{Hao.2020913,Chaccour.202051,Elzanaty.2021}, we only consider the LOS paths in the UE-RIS link and the AP-RIS link in the subTHz channels.
Different from most of the existing works, we consider the channel model with the near-field effects.
In the following, the formulation of signal propagation that accounts for the spherical curvature of the wavefront is provided.

\subsection{UE-RIS link}

We first consider the LoS path between UE $u$ and the RIS.
With the spherical wavefront, the receiving array response at the RIS's $m$-th element is given by 
\begin{equation}\label{ar1}
	\bm{a}_{R}[m] = \exp\left(-j \frac{2\pi}{\lambda} (d_{R,u}^m - d_{R,u}^0 ) \right),
\end{equation}
where $d_{R,u}^m$ represents the distance from the RIS's $m$-th element to UE $u$, and $d_{R,u}^0$ represents the distance from the central reflecting element to UE $u$.
Then, the complex channel gain of UE $u$-RIS link is denoted as 
\begin{equation}
	\bm{h}_{R,u} = g_{R,u}\bm{a}_{R}, 
\end{equation}
where $g_{R,u}$ represents the complex channel attenuation coefficient, and $\bm{a}_{R}$ collects the $N_R$ elements as $\bm{a}_{R} = [\bm{a}_{R}[1],\cdots,\bm{a}_{R}[N_R]]^T$.

Denote the center of the RIS panel as the phase reference point, and its coordinate is $(x_R,y_R,z_R)$.
Suppose that the $m$-th reflecting element is located at $(x_R,y_{R}+m_y\Delta_R ,z_{R}+m_z\Delta_R)$, where $m_y = -N^Y_{R}, \cdots, 0,N^Y_{R}$, and $m_z = -N^Z_{R}, \cdots, 0,N^Z_{R}$.
The index $m$ of the reflecting element can be expressed as
\begin{equation}\label{mmymz}
	m = m_z(2N^Y_{R}+1)+m_y,
\end{equation} 
and the value range of $m$ is given as $m = -N^Z_{R}(2N^Y_{R}+1)-N^Y_{R}, \cdots, N^Z_{R}(2N^Y_{R}+1)+N^Y_{R}$.

\begin{figure}
	\centering
	\vspace{-2em}
	\includegraphics[width=0.7\textwidth]{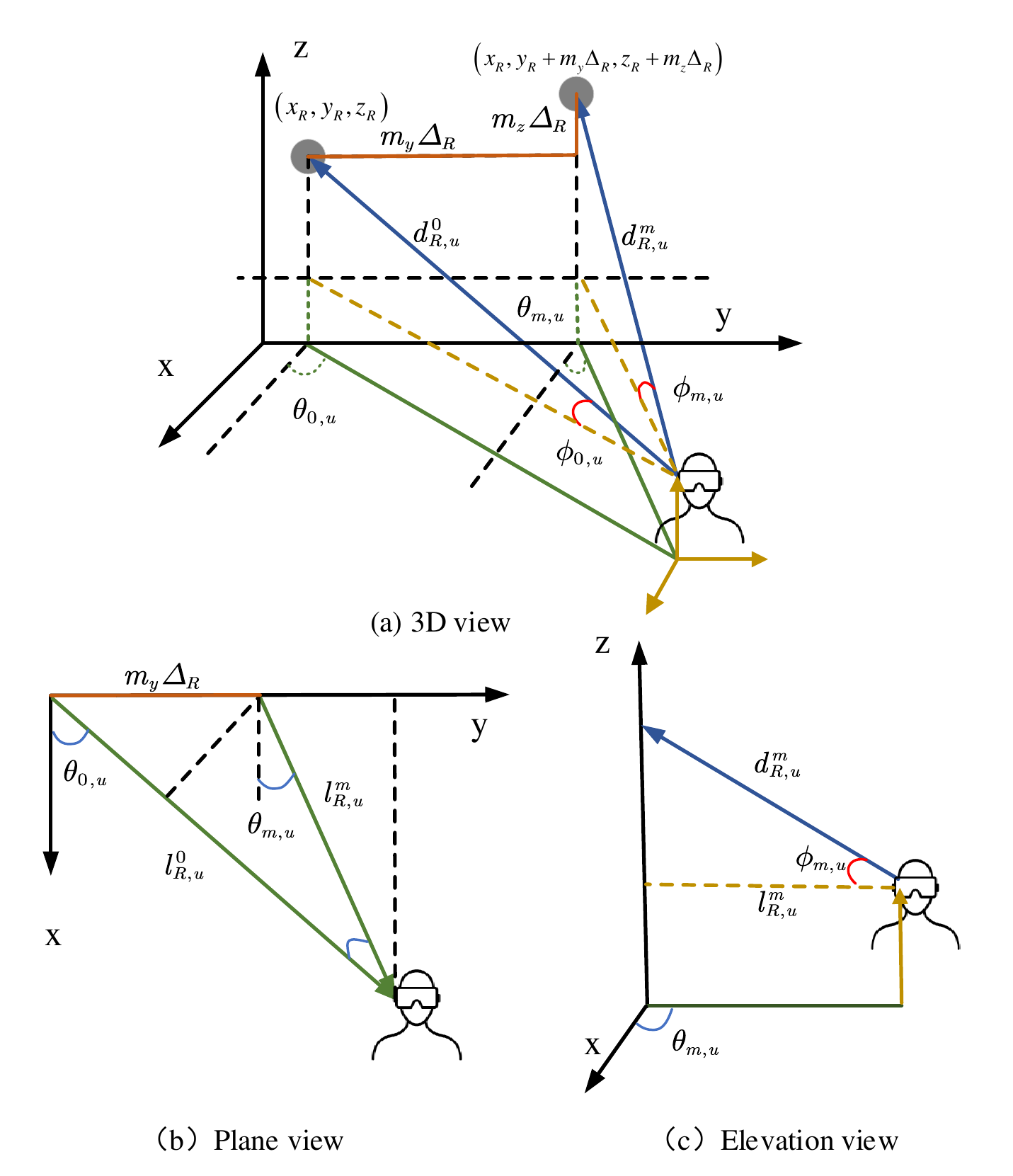}
	\vspace{-0.5em}
	\caption{Signal Propagation Model. }
	\vspace{-2em}
	\label{FigAnglesRIS}
\end{figure}

As shown in Fig. \ref{FigAnglesRIS},  the projection of distance $d_{R,u}^m$ on the XOY plane is  denoted as $l_{R,u}^m$, given by 
\begin{equation}\label{Lrum}
	l_{R,u}^m = \sqrt{(m_y\Delta_R)^2 +  (l_{R,u}^0)^2 - 2 \sin\theta_{u,0}m_y\Delta_R l_{R,u}^0},
\end{equation}
where $l_{R,u}^0$ denotes the projection of distance $d_{R,u}^0$ on the XOY plane, given by 
\begin{equation}\label{Lru0}
	l_{R,u}^0 = d_{R,u}^0\cos\phi_{u,0}.
\end{equation}

Then, the distance $d_{R,u}^m$ is calculated as 
\begin{equation} \label{Drum}
	d_{R,u}^m  =  \sqrt{(d_{R,u}^0\sin\phi_{u,0}+m_z\Delta_R)^2 + (l_{R,u}^m )^2}. 
\end{equation}

Substituting (\ref{Lrum}) and (\ref{Lru0}) into (\ref{Drum}), one obtains
\begin{equation}\label{dRsm}
	d_{R,u}^m =((m_y\Delta_R)^2 +  (m_z\Delta_R)^2 + (d_{R,u}^0)^2\\
	- 2d_{R,u}^0m_y\Delta_R\cos\phi_{u,0}\sin\theta_{u,0} + 2m_z\Delta_R d_{R,u}^0\sin\phi_{u,0} )^{\frac{1}{2}}.
\end{equation}


Letting $y = \frac{m_y\Delta_R}{d_{R,u}^0}$, and $z = \frac{m_z\Delta_R}{d_{R,u}^0}$,  the distance between the $m$-th reflecting element and UE $u$ can be represented as 
\begin{equation}\label{dRUm}
	d_{R,u}^m = d_{R,u}^0(y^2+z^2+1-2y\cos\phi_{u,0}\sin\theta_{u,0}\\
		+ 2z\sin\phi_{u,0})^{\frac{1}{2}} \triangleq d_{R,u}^0F(y,z). 
\end{equation}

In order to compute the distance $d_{R,u}^m$ in (\ref{dRUm}), we consider the following simplifications.
First of all, for ease of exposition, the first and second order derivatives of $F(y,z)$ with respect to y and/or z are given by
\begin{equation}
	\frac{\partial   F}{\partial y} = \frac{y-\cos\phi_{u,0}\sin\theta_{u,0}}{F(y,z)},
	\frac{\partial   F}{\partial z} = \frac{z+\sin\phi_{u,0}}{F(y,z)},
\end{equation}
\begin{align}
	\frac{\partial^2   F}{\partial y^2} = \frac{1+ z^2+2z\sin\theta_{u,0}-(\cos\phi_{u,0}\sin\theta_{u,0})^2}{F(y,z)^3}, \\
	\frac{\partial^2   F}{\partial z^2} =  \frac{1+ y^2-2y\cos\phi_{u,0}\sin\theta_{u,0}-(\sin\phi_{u,0})^2}{F(y,z)^3},\\
	\frac{\partial^2   F}{\partial z \partial y} = \frac{-(2y-2\cos\phi_{u,0}\sin\theta_{u,0})(z+\sin\phi_{u,0})}{2F(y,z)^3}.
\end{align}

\subsubsection{Far field approximation}

From \cite{selvan2017fraunhofer}, the far field of the RIS can be defined as the set of observation distances $R$ that are greater than the Fraunhofer distance $R_{f}$, i.e.,
\begin{equation}
	R\geq R_{f}=\frac{2L^{2}}{\lambda}\label{FDistance}.
\end{equation}
This definition corresponds to the maximum phase difference, $\pi/8$, of radio waves between any two reflecting elements.
In this paper, we select the center of the RIS panel as the reference point, as shown in (\ref{ar1}), thus the aperture of the RIS UPA should be the distance from panel vertex to the panel center as
\begin{equation}
	L = \sqrt{(M_y\Delta_R)^2+(M_z\Delta_R)^2}.\label{Lapt}
\end{equation}
As a result, when UE $u$ is located in the far field of the RIS panel, i.e., 
$$d_{R,u}^M >  2\frac{L^2}{\lambda} =  2\frac{(M_y\Delta_R)^2+(M_z\Delta_R)^2}{\lambda},$$ 
where $d_{R,u}^M$ denotes the distance of UE $u$ to the nearest RIS panel vertex.
Then, (\ref{dRsm}) can be approximated by the first-order Taylor expansion as
\begin{align}\label{dRsmLine} 
	d_{R,u}^m & \approx d_{R,u}^0 (F(y,z)|_{y=0,z=0}+ \frac{\partial   F}{\partial y}|_{y=0}(y) + \frac{\partial   F}{\partial z}|_{z=0}(z))  \nonumber\\
	& = d_{R,u}^0 (1 -\cos\phi_{u,0}\sin\theta_{u,0}y +\sin\phi_{u,0}z)  \nonumber\\
	& =  d_{R,u}^0 \left(\frac{m_z\Delta_R}{d_{R,u}^0}\sin\phi_{u,0} -\frac{m_y\Delta_R}{d_{R,u}^0}\sin\theta_{u,0}\cos\phi_{u,0}+ 1\right)\nonumber\\
	& = m_z\Delta_R\sin\phi_{u,0} -m_y\Delta_R\sin\theta_{u,0}\cos\phi_{u,0}+ d_{R,u}^0.
\end{align}

Therefore, in this case, the array response is characterized only by the azimuth and vertical AoAs. 
For ease of exposure, we use $\omega_u = \sin\phi_{u,0}$ and $\varphi_u = \sin\theta_{u,0}\cos\phi_{u,0}$ to denote the AoAs.
Then, the array response for the far field can be denoted as 
\begin{equation}\label{aF}
	\bm{a}^F_{R}(\omega_u,\varphi_u)[m]  =  \exp\left(-j \frac{2\pi}{\lambda} (m_z\Delta_R\omega_u -m_y\Delta_R\varphi_u) \right).
\end{equation}
where the relationship between index $m$ and $(m_y,m_z)$ was given by (\ref{mmymz}).

According to (\ref{FDistance}), the Fraunhofer distance would be very large if an extra-large RIS is utilized.
If the RIS is working at high-frequency bands, such as sub-THz,  the Fraunhofer distance also increases due to the smaller wavelength \cite{RenzoRIS2020}. 
Therefore, it is very likely that some UEs are in the near field when the RIS is deployed for indoor applications such as localization.
In this case, the spherical wavefront of the EM waves should be considered.

\subsubsection{Near field formulation}

When UE $u$ is located at the near field of the RIS panel, i.e., the Fresnel region 
$$d_{R,u}^m \in \left[0.62\frac{L^3}{\sqrt{\lambda}}, 2\frac{L^2}{\lambda}\right],$$
a good approximation is based on the Fresnel approximation, which corresponds to the second-order Taylor expansion\cite{Zheng.2019}.
So far, the second-order Fresnel approximation results have been developed for the array with ULA geometry\cite{Zheng.2019,Tao.2011,Zuo.2018}, but they cannot be directly applied to the two-dimensional RIS UPA case.
Thus, we derive the two-dimensional Fresnel approximation as
\begin{equation}\label{Fyz}
	F(y,z) \approx  F(y,z)|_{y=0,z=0}+ \frac{\partial   F}{\partial y}|_{y=0}\times y + \frac{\partial   F}{\partial z}|_{z=0} \times z\\
	 + \frac{1}{2}[y,z]\times\Delta^2\bm{F}\times[y,z]^T,
\end{equation}
where 
\begin{equation}
\begin{split}
\Delta^2\bm{F} &= \begin{bmatrix} 	\frac{\partial^2   F}{\partial y^2}|_{y=0,z=0}  & \frac{\partial^2   F}{\partial z \partial y}|_{y=0,z=0} \\ 	\frac{\partial^2 F}{\partial z \partial y}|_{y=0,z=0} & \frac{\partial^2   F}{\partial z^2}|_{y=0,z=0} \end{bmatrix} \\
&=\begin{bmatrix} 1-(\cos\phi_{u,0}\sin\theta_{u,0})^2  & \cos\phi_{u,0}\sin\theta_{u,0}\sin\phi_{u,0}  \\ 	\cos\phi_{u,0}\sin\theta_{u,0}\sin\phi_{u,0} &1 -(\sin\phi_{u,0})^2 \end{bmatrix}. 
\end{split}\nonumber
\end{equation}

Then, (\ref{Fyz}) can be reformulated as 
\begin{equation}\label{FyzFres}
	F(y,z) \approx 1 + \omega_u z -\varphi_u y + \frac{1}{2}(\overline{\omega}_u y^2 + \overline{\varphi}_u z^2 +2\varphi_u \omega_u yz ),
\end{equation}
where $\overline{\omega}_u = 1- \omega_u^2$, and $\overline{\varphi}_u= 1- \varphi_u^2$.
Then,  the distance of the $m$-th reflecting element $d_{R,u}^m$ can be approximated as 
\begin{align}
	\!\!\!\!\!\!\!\quad d_{R,u}^m & \approx  m_z\Delta_R\omega_u -m_y\Delta_R\varphi_u  + \frac{1}{2}\overline{\omega}_u\frac{(m_y\Delta_R)^2}{d_{R,u}^0}
	  +\frac{1}{2} \overline{\varphi}_u \frac{(m_z\Delta_R)^2}{d_{R,u}^0} +  \varphi_u \omega_u\frac{m_y\Delta_R(m_z\Delta_R)}{d_{R,u}^0} + d_{R,u}^0 \nonumber\\
	&= m_z\Delta_R\omega_u -m_y\Delta_R\varphi_u  - \frac{\left(m_z\Delta_R\omega_u - m_y\Delta_R\varphi_u \right)^2}{2d_{R,u}^0}  + \frac{1}{2d_{R,u}^0}\left( (m_z\Delta_R)^2 +(m_y\Delta_R)^2\right) + d_{R,u}^0 \nonumber\\
	&\triangleq J_{m}(\omega_u,{\varphi_u }) + Q_{m}(\omega_u,{\varphi_u },d^0_{R,u}) + d_{R,u}^0,     \label{DeltaDR}
\end{align}
where $J_{m}(\omega_u,{\varphi_u }) =  m_z\Delta_R\omega_u -m_y\Delta_R\varphi_u$, and
\begin{equation}
\begin{split}
Q_{m}(\omega_u,{\varphi_u },d^0_{R,u})  = \frac{1}{2d^0_{R,u}}\left( (m_z\Delta_R)^2 +(m_y\Delta_R)^2\right) - \frac{1}{2d^0_{R,u}}\left(m_z\Delta_R\omega_u - m_y\Delta_R\varphi_u \right)^2.\end{split} \nonumber
\end{equation}

In the near field case, apart from the azimuth and vertical angles, the array response  is also dependent on the distance $d_{R,u}^0$, which is the distance from UE $u$ to the center of RIS.
Then, the  array response for the near field UE can be approximated as 
\begin{equation}\label{ardm}
	\bm{a}^N_{R}(\omega_u,\varphi_u,d_{R,u}^0)[m]  =  \exp\left(-j \frac{2\pi}{\lambda}(J_{m}(\omega_u,{\varphi_u })  + Q_{m}(\omega_u,{\varphi_u },d_{R,u}^0)) \right).
\end{equation}
It can be verified that $\bm{a}_{R}(\omega_u,\varphi_u ,d_{R,u}^0)[m] \to  \bm{a}_{R}(\omega_u,\varphi_u )[m] $ when $d_{R,u}^0$ is sufficiently large.
In other words, the near field case reduces to the far-field case.
Then, in the near field, the complex channel gain from UE $u$ to the RIS is denoted as 
\begin{equation}
	\bm{h}_{R,u} =  g_{R,u} \bm{a}^N_{R}(\omega_u,\varphi_u,d_{R,u}^0),
\end{equation}
where the $m$-th element of $\bm{a}^N_{R}(\omega_u,\varphi_u,d_{R,u}^0)$ is given by (\ref{ardm}).

\begin{figure}
	\centering
	\vspace{-2em}
	\begin{subfigure}{0.45\textwidth}
		\includegraphics[width=\textwidth]{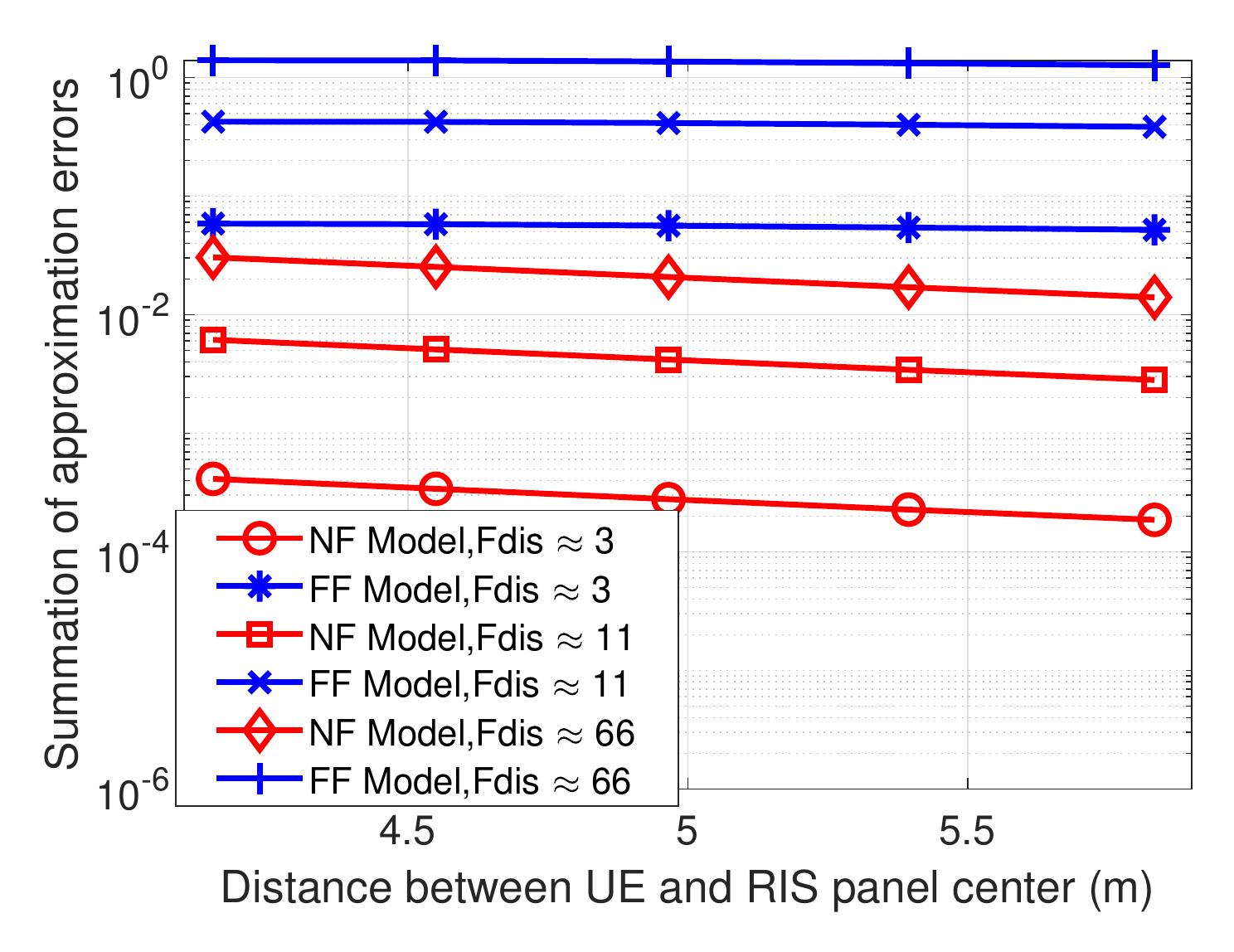}
		\subcaption{Approximation errors of distances.}
		\label{FigErrRIS_Dis}
	\end{subfigure}
	\hspace*{\fill} 
	\begin{subfigure}{0.45\textwidth}
		\includegraphics[width=\textwidth]{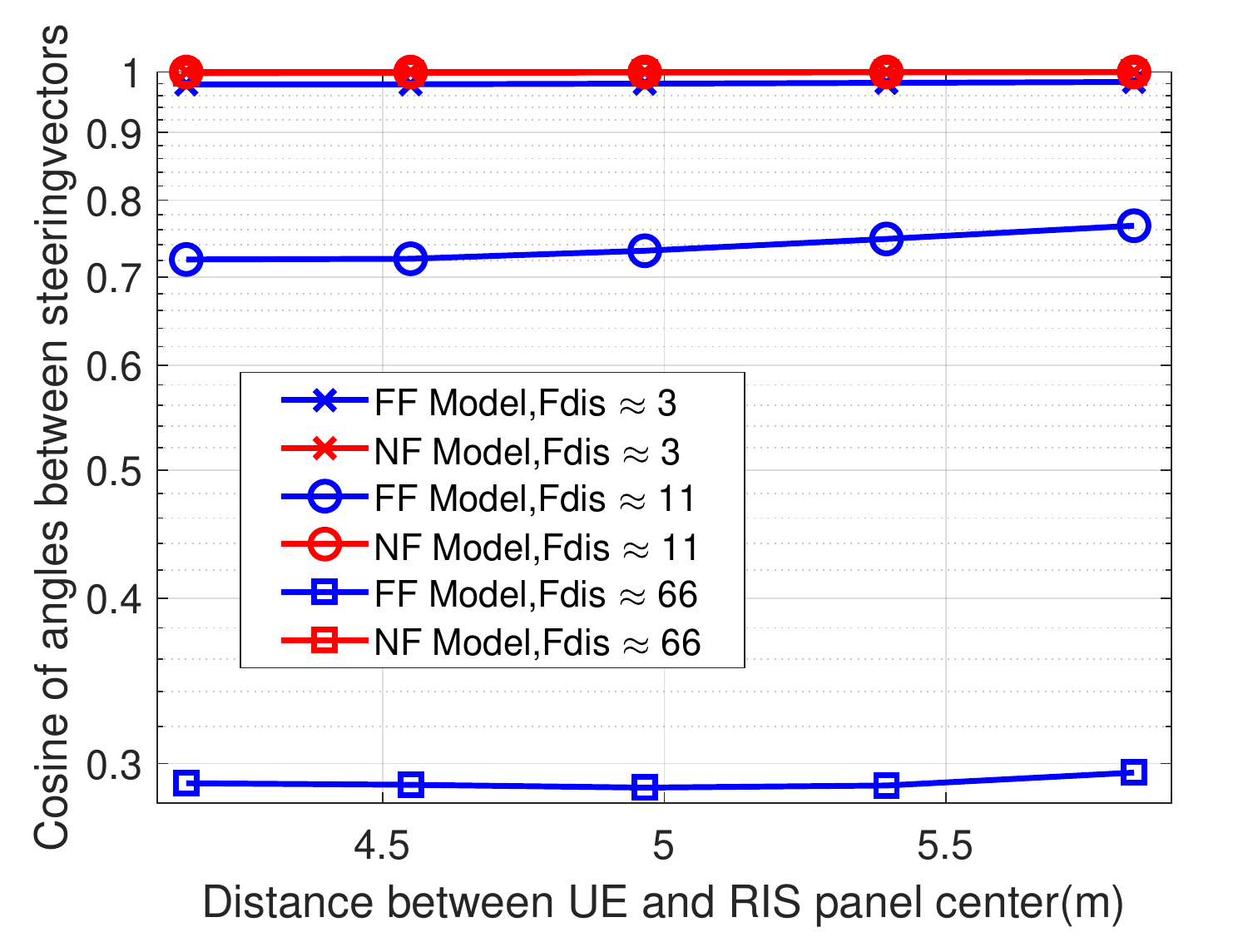}
		\subcaption{Cosine of the angles between steering vectors. }
		\label{FigErrRIS_Element}
	\end{subfigure}
	\caption{Performance of  near/far-field approximation.}
	\label{FigErrRIS}
	\vspace{-2em}
\end{figure}

Overall, the distance $d_{R,u}^m$ can be approximated by the near-field model (\ref{DeltaDR}) and the far-field model (\ref{dRsmLine}), and the approximation errors are shown in Fig. (\ref{FigErrRIS_Dis}) with labels ``NF'' and ``FF'', respectively.
In  Fig. (\ref{FigErrRIS_Dis}), the Y-axis denotes the summation of approximation errors of $d_{R,u}^m$ for all $m = -N_R^Z(N_R^Y+1)-N_R^Y, \cdots,N_R^Z(N_R^Y+1)+N_R^Y$.
As expected, it is observed that the proposed near-field model is more accurate than the far field model, 
as the approximation errors of ``NF'' are much less than that of ``FF''.
Also, when the distance between UE and the RIS panel increases, the approximation error of the near-field model also decreases, which indicates that the ``NF'' model becomes more accurate for longer transmission distance.
Also, the array response given in (\ref{ar1}) can be approximated by the near-field steering vector (\ref{ardm}) and the far-field steering vector (\ref{aF}), respectively, i.e., $\bm{a}_{R}\approx \bm{a}^N_{R}$ and $\bm{a}_{R}\approx \bm{a}^F_{R}$.
Fig. (\ref{FigErrRIS_Element}) shows the cosine of the angles between $\bm{a}_{R}$ and  $\bm{a}^N_{R}$, and the cosine of the angles between $\bm{a}^F_{R}$ and $\bm{a}_{R}$\footnote{The simulation configurations of Fig. (\ref{FigErrRIS_Dis}) and Fig. (\ref{FigErrRIS_Element}) as the same as that in Section \ref{sumSec}, where the details of configurations can be found.}.
It is observed that the near-field $\bm{a}^N_{R}$ is very close to $\bm{a}_{R}$, i.e. the value of cosine is equal to $1$, while the approximation error of $\bm{a}^F_{R}$ is much larger.
In addition, the cosine values of the $\bm{a}^F_{R}$ slightly increases with the distance as the near-field effects decreases.

\subsection{ RIS-AP link} 

It is assumed that the LoS channel between the RIS and the AP is varying slowly.
Due to the spherical nature of the wavefront, the transmission distances from each element of the RIS to each antenna element of the AP array determine the phases of the received signals at the AP.

According to \cite{Bohagen.2007,Bohagen.2009,bohagen2007optimal}, a high-rank channel model of the RIS-AP link can be obtained with a proper deployment.
With the spherical wave modeling, the normalized channel gain of the RIS-AP link is denoted as $\bm{G}_{A,R}$, which collects the channel elements of the RIS-AP LoS link.
The $(m,n)$-th element of the $\bm{G}_{A,R}$ is given by \cite{Bohagen.2007} 
\begin{equation}
	\bm{G}_{A,R}(m,n) =  \exp\left(j \frac{2\pi}{\lambda} r_{m,n} \right),
\end{equation}
where $r_{m,n}$ denotes the path length between the $m$th antenna element of the AP and the $n$-th reflecting element.
Letting $g_{A}$ denote the common large-scale path loss attenuation, the complex channel gain is then given by\cite{Bohagen.2009}  
\begin{equation}
	\bm{H}_{A,R} = g_{A} \bm{G}_{A,R}.
\end{equation}

\subsection{Cascaded Channel}

Let $\bm{E}_{RIS} = \text{diag}\{\bm{e}\} \in \mathbb{C}^{N_R\times 1}$ denote the phase shift matrix of the RIS,  where $\bm{e}$ denotes the phase shift vector of the RIS.
Then, the RIS-assisted channel of UE $u$ is expressed as
\begin{equation}
	\bm{h}_u = g_{R,u}\sqrt{P_{u}^t}\bm{H}_{A,R}\bm{E}_{RIS}\bm{a}_{R,u} = g_{u}\sqrt{P_{u}^t}\bm{G}_{A,R}\bm{E}_{RIS}\bm{a}_{R,u} ,
\end{equation}
where $\sqrt{P_{u}^t}$ denotes the transmit power of UE $u$, $g_{u}$ is the cascaded complex channel attenuation defined as $g_{u} = g_{A}g_{R,u}$, and $\bm{a}_{R,u}  = \bm{a}^N_{R}(\omega_u,\varphi_u,d_{R,u}^0)$.

In time slot $t$, the transmit signal vector of all the $U$ UEs is denoted as $\bm{x}_t= [x_1(t),\cdots,x_U(t)]^T$ $ \in \mathbb{C}^{U\times 1}$.
With the assistance of the RIS, the received signal at the AP is modeled as 
\begin{align}
	\bm{y}(t) &= \sqrt{P_{u}^t}\bm{G}_{A,R}\bm{E}_{RIS}(t)\sum_{u=1}^U  g_{u}\bm{a}_{R,u}x_u(t) + \bm{n}(t) \nonumber\\
	&= \sqrt{P_{u}^t}\bm{G}_{A,R}\bm{E}_{RIS}(t)\bm{A}\bm{x}_t + \bm{n}(t),\label{yt}
\end{align} 
where  $\bm{A} = [g_{1}\bm{a}_{R,1}, \cdots,g_{u} \bm{a}_{R,u}, \cdots,g_{U} \bm{a}_{R,U}]$, and 
$\bm{n}(t)$ denotes the Gaussian noise vector with the distribution of $\mathcal{CN}(0,\sigma^2\bm{I})$.

In the AP-RIS link, as the locations of the RIS and the AP are known as prior, the distances $\{r_{m,n}\}$ between the elements of the AP and that of the RIS can be readily obtained.
First, the unknown channel matrix $\bm{A}$ should be estimated.
Then, to locate UE $u$, the unknown parameters in $\bm{A}$ that need to be estimated are
\begin{itemize}
	\item  $\omega_u$, $\varphi_u $ and $d_{R,u}^0$, which depends on the location of UE $u$ in the RIS-UE $u$ link.
	\item  $g_u$, which is  the complex channel attenuation coefficient.
\end{itemize}

\section{Joint Channel and Localization Estimation}

\subsection{RIS Training Phase Shifts and Pilot Design}

First, we need to obtain the unknown channel matrix $\bm{A}$.
The simplest method is to adopt the LS method.
However, in the considered scenario, the number of antenna elements of the AP is less than that of the RIS panel, i.e., $N_A < N_R$, resulting in that $\bm{G}_{A,R} \in \mathbb{C}^{N_A \times N_R}$ has the rank of $N_A$.
Then, the LS estimation cannot be directly applied to (\ref{yt}).
Therefore, to obtain the unique estimation of $\bm{A}$, we utilize different RIS training phase shift vectors $\bm{e}(t)$ and pilot data $\bm{x}_t$.

\begin{figure}
	\centering
	\vspace{-2em}
	\includegraphics[width=0.4\textwidth]{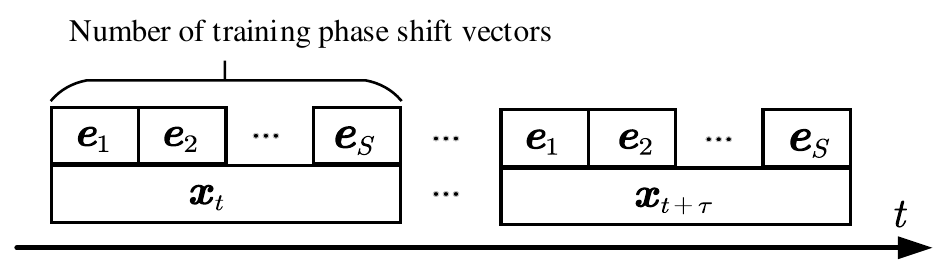}
	\vspace{-0.5em}
	\caption{Design of RIS phase shift vectors and the pilot data.}
	\vspace{-1em}
	\label{FigPilot}
\end{figure}

Fig. \ref{FigPilot} shows the numbers of required RIS training phase shift vectors and pilot data sets.
As shown in Fig. \ref{FigPilot}, the RIS phase shift vector changes for $S$ times while the pilot data keeps fixed, and the set of phase shift vectors is denoted as $\{\bm{e}_1,\cdots,\bm{e}_S\}$.
The number of RIS training phase shift vectors is denoted as $S$, and the required number of pilot data is denoted as $\tau$.
Let $\bm{y}_t$ denote the composited signal for the $t$-th pilot data duration, which collects the received signal with $S$ different  phase shift vectors as
\begin{equation}
	\bm{y}_t =  \bm{G}_{RIS}\bm{A}\bm{x}_t+ \bm{n}_t,
\end{equation}
where the received vector $\bm{y}_t $ has the size of $N_AS\times 1$,
$\bm{y}_t = [\bm{y}(t_1)^T,\cdots,\bm{y}(t_S)^T]^T$, and
$\bm{n}_t = [\bm{n}(t_1)^T,\cdots,\bm{n}(t_S)^T]^T$.
The matrix $\bm{G}_{RIS}$ with size $N_AS\times N_R$ collects the composite channels introduced by $S$ different phase shift vectors, which are $\bm{G}_{RIS} = [ (\bm{G}_{A,R}\text{diag}{(\bm{e}_1)})^T, \bm{G}_{A,R}\text{diag}{(\bm{e}_2)}^T, \cdots,\\ \bm{G}_{A,R}\text{diag}{(\bm{e}_S)}^T]^T$.

Then, we have the LS estimation for $\bm{A}\bm{x}_t$ as 
\begin{equation}\label{AA}
	\widehat{\bm{A}\bm{x}}_t = (\bm{G}_{RIS} ^H\bm{G}_{RIS} )^{-1} \bm{G}_{RIS} ^H\bm{y}_t = \bm{H}^{\dag}_{RIS}\bm{y}_t.
\end{equation}

\textit{Remark 1}: 
It is observed that the rank of cascaded channel $\bm{G}_{RIS}$ can be increased by applying different RIS phase shifts, so that the Moore-Penrose pseudoinverse of $\bm{G}_{RIS}$ denoted as $\bm{H}^{\dag}_{RIS}$ can be obtained.
As demonstrated in \cite{Kundu}, the Discrete Fourier Transform (DFT) RIS phase training scheme can achieve the nearly optimal performance in terms of the mean squared error (MSE). 
Thus, the training phase shift matrix ${\bm{E}_S=[\bm{e}_1,\bm{e}_2,\cdots,\bm{e}_S]}$ can be set as the first $S$ columns of the $N_R \times N_R$ DFT matrix.

For transmit pilot design, we assume that transmit signal from UE $u$ is denoted as $x_u$ following the distribution of $\mathbb{E}[|x_u|^2] =1$ and $\mathbb{E}[x_u x_k] = 0$, for all $k \neq u$. 
Consequently, the array covariance matrix can be estimated as 
\begin{equation}\label{Rnoise}
	\tilde{\bm{R}} =  \mathbb{E}[\bm{H}^{\dag}_{IRS}\bm{y}_t \bm{y}_t^H(\bm{H}^{\dag}_{IRS})^H] 
	=\bm{A}\bm{A}^H+ \sigma^2(\bm{G}_{RIS} ^H\bm{G}_{RIS})^{-1}.
\end{equation}

By collecting the received signal $\bm{y}_t$ with different $\tau$ pilot durations, the estimated array covariance matrix $\hat{\bm{R}}$ of $\tilde{\bm{R}}$ is given by 
\begin{equation} \label{Rhat}
	\hat{\bm{R}} = \frac{1}{\tau} \sum_{t=1}^{\tau} \widehat{\bm{A}\bm{x}}_t (\widehat{\bm{A}\bm{x}}_t)^H,
\end{equation}
where the total number of snapshots is $\tau$.

\textit{Remark 2}: 
In order to ensure that the matrix $\bm{G}_{RIS}$ is full column rank, the number of training phase shift vectors should satisfy $S > \frac{N_R}{N_A}$.
Moreover, the output noise of the LS estimator is amplified by $(\bm{G}_{RIS} ^H\bm{G}_{RIS})^{-1}$ as shown in (\ref{Rnoise}).
Theoretically, the Hermitian matrix $\bm{G}_H = \bm{G}_{RIS} ^H\bm{G}_{RIS}$ may have rank deficiency, i.e.  the minimum singular values of $\bm{G}_H$ is sufficiently small so that $\rho_{min}(\bm{G}_H) \to 0$.
In this case, the noise will be significantly amplified.
Therefore, the singular values of the constructed Hermitian matrix should be large enough.
Thus, in order to suppress noise, the number of RIS training phase shift vectors should be sufficiently large, and its impact on the performance will be give in Section \ref{sumSec}.

\subsection{Estimation of AoAs}

Without noise, we have the following explicit expression
\begin{align}
	\bm{R} \!\!=\!\!\bm{A}\bm{A}^H \!\!\!= \sum_{u =1}^U  |g_{R,u}|^2 \bm{a}_{R,u}\bm{a}^H_{R,u}.
\end{align}

Then, the $(p,q)$-th element of covariance $\bm{R}$ is given by 
\begin{equation}
	\bm{R}(p,q) = \sum_{u =1}^U  |g_{R,u}|^2 \exp\left(-j \frac{2\pi}{\lambda}(J_{p,q} + Q_{p,q}) \right), 
\end{equation}
where $J_{p,q} = J_{p}(\omega_u,{\varphi_u })  -J_{q}(\omega_u,{\varphi_u })$,
$Q_{p,q}= Q_{p}(\omega_u,{\varphi_u },d_{R,u}^0) - Q_{q}(\omega_u,{\varphi_u },d_{R,u}^0)$.
To be specific, 
\begin{align}
Q_{p,q}&=  \frac{1}{2d_{R,u}^0}\left( (p_z\Delta_R)^2 +(p_y\Delta_R)^2\right)- \frac{1}{2d_{R,u}^0}\left(p_z\Delta_R\omega_u - p_y\Delta_R\varphi_u \right)^2 \nonumber\\
	& - \frac{1}{2d_{R,u}^0}\left( (q_z\Delta_R)^2 +(q_y\Delta_R)^2\right) + \frac{1}{2d_{R,u}^0}\left(q_z\Delta_R\omega_u - q_y\Delta_R\varphi_u \right)^2 . \nonumber\\
J_{p,q}&=  (p_z-q_z)\Delta_R\omega_u -(p_y-q_y)\Delta_R\varphi_u , \nonumber
\end{align}
where indices are given by
$p(p_z,p_y) = N^Z_{R}(2N^Y_{R}+1)+ N^Y_{R} + p_z(2N^Y_{R}+1)+p_y$,
$q(q_z,q_y) = N^Z_{R}(2N^Y_{R}+1)+ N^Y_{R} + q_z(2N^Y_{R}+1)+q_y$,
$ p_y,q_y = -N^Y_{R}, \cdots, N^Y_{R},p_z,q_z = -N^Z_{R}, \cdots, N^Z_{R}$.
Then, it is found that for all  $p_y = - q_y$, and $p_z = - q_z$, one obtains
$$J_{p,q} =  2p_z\Delta_R\omega_u - 2p_y\Delta_R\varphi_u, \qquad Q_{p,q} =0 .$$
In other words, the quadratic term is eliminated for the element of $\bm{R}(p(p_z,p_y),q(-p_z,-p_y))$ in the covariance matrix $\bm{R}$, which are 
\begin{equation}
\bm{R}(p(p_z,p_y),q(-p_z,-p_y)) 
	= \sum_{u =1}^U\!|g_{R,u}|^2\!\exp\!\left(\!-j \frac{2\pi}{\lambda}(2J_{p}(\omega_u,{\varphi_u }) )\!\right).\nonumber
\end{equation}

We provide the following approach to separate the distance and the AoAs in the covariance matrix.  
First, we define steering vectors as 
\begin{align}
&\bm{v}_{u}(\omega_u)[z] =  \exp\left(-j \frac{2\pi}{\lambda}(2z\Delta_R\omega_u)\right), z= 0, \cdots,N^Z_{R}, \nonumber \\
&\bm{s}_{u}(\varphi_u)[y] =  \exp\left(-j \frac{2\pi}{\lambda}(-2y\Delta_R\varphi_u)\right),y = 0, \cdots,N^Y_{R}, \nonumber \\
& \bm{b}_{u}(\omega_u, \varphi_u) =  \bm{v}_{u}(\omega_u) \otimes \bm{s}_{u}(\varphi_u)  = [\bm{v}_{u}(\omega_u)[0]\bm{s}^T_{u}(\varphi_u), \cdots, \bm{v}_{u}(\omega_u)[N^Z_{R}]\bm{s}^T_{u}(\varphi_u)]^T.\nonumber
\end{align}
Then, we define the correlation matrix $\bm{R}_b$ as  
\begin{equation}
	\bm{R}_b= \sum_{u =1}^U  |g_{R,u}|^2 \bm{b}_{u}(\omega_u, \varphi_u)\bm{b}^H_{u}(\omega_u, \varphi_u),
\end{equation}
Defining the following function
\begin{equation}
	e(z,y) = \sum_{u =1}^U  |g_{R,u}|^2 \exp\left(-j \frac{4\pi}{\lambda}(z\Delta_R\omega_u -y\Delta_R\varphi_u ) \right),
\end{equation}
we can represent the covariance matrix $\bm{R}_b$ as (\ref{eqn_dbl_x}).
Note that 
\begin{equation}
	\bm{R}(p(z,y) ,q(-z,-y)) =  e(z,y) = (e(-z,-y))^*. \nonumber
\end{equation}
For instance, the first element of the matrix $\bm{R}_b$, i.e., $e(0,0)$ is the same as $\bm{R}(p(0,0) ,q(0,0))$, where
$p(0,0) = N^Z_{R}(2N^Y_{R}+1)+ N^Y_{R} = \frac{N_R-1}{2},q(0,0) = N^Z_{R}(2N^Y_{R}+1)+ N^Y_{R} = \frac{N_R-1}{2}$.

\begin{figure*}[!t]
	\vspace*{-1em}
	\normalsize
	\setcounter{MYtempeqncnt}{\value{equation}}
	\setcounter{equation}{33}
\begin{equation}\label{eqn_dbl_x}
	\bm{R}_b =\begin{bmatrix} 	
		e(0,0) &  \! e(0,-1) &  \cdots  & e(0,-N^Y_{R}) \!& \!\!\!\!\!\! e(-1,0) &\!\! \cdots \!\!& e(-N^Z_{R},-N^Y_{R})\\
		e(0,1) &  \! e(0,0) &  \cdots  & e(0,-N^Y_{R}+1) \!& \!\!\!\!\!\!   e(-1,1) & \!\! \cdots \!\! & e(-N^Z_{R},-N^Y_{R}+1)\\
		\vdots & \! \vdots &  \ddots  &   \vdots           \!& \!\!\!\!\!\!  \vdots  \!& \!\!   \ddots\!\!&\vdots\\
		e(0,N^Y_{R}) &  \! e(0,N^Y_{R}-1) &  \cdots  \!& \!\! e(0,0) \!& \!\!\!\!\!\!   e(-1,N^Y_{R}) \!\!&  \cdots  & e(-N^Z_{R},0)\\
		\vdots & \vdots &  \! \ddots  &   \vdots           &\!\!\!\!\!\! \vdots  \!& \!\!   \ddots  \!\!&\vdots\\
		e(N^Z_{R},N^Y_{R}) &  \! e(N^Z_{R},N^Y_{R}-1)\!& \!\! \cdots \!& \!\!  e(N^Z_{R},0) \!& \!\!\!\!\!\!  e(N^Z_{R}-1,N^Y_{R}) & \!\! \cdots\!\!  & e(0,0)\\
	\end{bmatrix}.
\end{equation}
\begin{equation}\label{eqn_dbl_y}
		\bm{S}_{i}\!= \!\!\!\begin{bmatrix} 	
			\bm{R}(N_i,\overline{N}_{i})  &  \bm{R}(N_i-1,\overline{N}_{i}+1) &\!\!\! \cdots \!\!\!&\!\!\! \bm{R}(N_i-N^Y_{R},\overline{N}_{i} +N^Y_{R})  \\
			\bm{R}(N_i+1,\overline{N}_{i}-1)  &  \bm{R}(N_i,\overline{N}_{i}) &\!\!\! \cdots \!\!\!&\!\!\! \bm{R}(N_i-N^Y_{R}-1,\overline{N}_{i} +N^Y_{R}+1)  \\
			\vdots&\!\!\! \vdots\!\!\!&\!\!\!  \ddots  & \vdots \\
			\bm{R}(N_i + N^Y_{R},\overline{N}_{i}-N^Y_{R})  &  \bm{R}(N_i + N^Y_{R}-1,\overline{N}_{i}-N^Y_{R}+1) & \!\!\!\cdots \!\!\!&\!\!\! \bm{R}(N_i,\overline{N}_{i})  \\
		\end{bmatrix}.
\end{equation}
	\setcounter{equation}{\value{MYtempeqncnt}}
	\hrulefill
\end{figure*}

Let $N_b$ represent the length of vector $\bm{b}_u$, and $N_b = (N^Z_{R} +1) \times (N^Y_{R} +1)$.
The index of the center reflecting element is denoted as $N_0$, and $N_0 = \frac{N_R-1}{2}$.
Then, the index functions can be represented as 
$p(z,y) = N_0 + z(2N^Y_{R}+1)+y$, and $q(-z,-y) = N_0 - z(2N^Y_{R}+1)-y$.
Letting $N_{i} = N_0+ i(2N^Y_{R} +1) $ and $\overline{N}_{i} = N_0 - i(2N^Y_{R} +1)$, the element matrix $\bm{S}_{i}$ with size  $(N^Y_{R}+1) \times (N^Y_{R}+1)$ is given by (\ref{eqn_dbl_y}).

\addtocounter{equation}{2}
Utilizing $\bm{T}_{i}$, we construct the down-sampled Toeplitz matrix with size $ N_b\times  N_b $ as 
\begin{equation} \label{ToepMat}
	\bm{T}=\begin{bmatrix} 	
		\bm{S}_0  &  \bm{S}_{-1}  & \cdots   &  \bm{S}_{-N^Z_{R}}  \\
		\bm{S}_1  &  \bm{S}_{0}   &  \cdots   &  \bm{S}_{-N^Z_{R}+1}  \\
		\vdots       &   \vdots           &  \ddots  &  \vdots \\
		\bm{S}_{N^Z_{R}}  &  \bm{S}_{N^Z_{R}-1}   &  \cdots   & \bm{S}_{0}  \\
	\end{bmatrix}.
\end{equation}

We can derive that 
\begin{equation}
	\bm{T}=  \sum_{u =1}^U  |g_{R,u}|^2 \bm{b}_{u}(\omega_u, \varphi_u)\bm{b}^H_{u}(\omega_u, \varphi_u).
\end{equation}
It is observed that the sampled correlation matrix $\bm{T}$ only depends on the azimuth and vertical angles, and which is the same as the far-field case.

\textit{Remark 3}: 
Through this method, the distance and the AoAs can be decoupled by the sampled correlation matrix $\bm{T}$.
Then, the conventional far-field angle estimation approaches can be leveraged in this case, such as the two-dimensional MUSIC-like spectrum peak searching\cite{Liang.2010}.
However, note that the MUSIC algorithm requires eigendecomposition.
In the following, we propose a computationally efficient subspace-based method for the AoA estimation.

First, we define matrix $\bm{B} = [g_{R,1}\bm{b}_{1},\cdots,g_{R,U}\bm{b}_{U}]$, and divide the $N_b\times U$ matrix into two parts as 
\begin{equation}\label{B1B2}
	\bm{B} \triangleq  \left[\begin{array}{c} 
		\bm{B}_1\in  \mathbb{C}^{\small{U} \times \small{U}} \\ 
		\bm{B}_2\in  \mathbb{C}^{ (\small{N_R -U}) \times \small{U}} 
	\end{array}\right]
\end{equation}
where matrix $\bm{B}_1$ contains the first $U$ rows of $\bm{B}$ and $\bm{B}_2$ contains the remaining $N_b -U$ rows.
Note that $\bm{B}_1$ is a Vandermonde matrix.
Assuming that the $U$ UEs are from distinct directions, $\bm{B}_1$ has the full rank of $U$, and the rows of $\bm{B}_2$ can be expressed as a linear combination of linearly independent rows of $\bm{B}_1$\cite{Xin.2004b,MARCOS1995121}. 
Equivalently, there is a $U \times (N_b -U)$ linear operator $\bm{P}_1$ between  $\bm{B}_1$ and $\bm{B}_2$ as 
\begin{equation}
	\bm{P}_1^H \bm{B}_1 = \bm{B}_2,
\end{equation}
and the linear operator $\bm{P}_1$ can be calculated as 
\begin{equation}\label{P1B1}
	\bm{P}_1 = (\bm{B}_1^H)^{-1} \bm{B}_2^H.
\end{equation}
Then, we utilize the correlation matrix to obtained linear operator $\bm{P}_1$.
The correlation matrix $\bm{T}$ can be represented as 
\begin{equation}\label{T1T2}
	\bm{T}  = \bm{B}\bm{B}^H =\begin{bmatrix} \bm{B}_1\bm{B}^H \\ \bm{B}_2\bm{B}^H\end{bmatrix}  =\begin{bmatrix} \bm{T}_1  \in  \mathbb{C}^{U\times N_R}\\ \bm{T}_2 \in \mathbb{C}^{(N_R-U) \times N_R} \end{bmatrix} \begin{matrix}  \end{matrix}
\end{equation}
where $\bm{T}_1$ consists of the first $U$ rows and $\bm{T}_2$ consists of the remaining $(N_b- U)$ rows.
Then, the relationship between the submatrices of $\bm{T}_1$ and $\bm{T}_2$ is 
\begin{equation}\label{PT1T2}
	\bm{P}_1^H \bm{T}_1 = \bm{T}_2.
\end{equation}
Then, the linear operator $\bm{P}_1$ can be found from $\bm{T}_1$ and $\bm{T}_2$ as
\begin{equation}\label{POp}
	\bm{P}_1 = (\bm{T}_1\bm{T}_1^H)^{-1}\bm{T}_1 \bm{T}_2^H.
\end{equation}

With the obtained linear operator $\bm{P}_1$, we have 
\begin{equation}
	\bm{Q}^H\bm{B}  = \bm{0}_{(N_b- U) \times N_b},
\end{equation}
where  
$\bm{Q}^H = [\bm{P}_1^H, -\bm{I}_{N_b- U}]$.
As a result, the columns of $\bm{Q}$ are in the null space of $\bm{B}$, i.e. $\mathcal{N}(\bm{B})$ and the orthogonal projector onto this subspace $\mathcal{N}(\bm{B})$ is given by 
\begin{equation}\label{PiQ}
	\bm{\Pi}_Q \triangleq  \bm{Q} (\bm{Q}^H \bm{Q})^{-1}\bm{Q}^H .
\end{equation}
Then, we can infer that $\bm{\Pi}_Q \bm{b}_u = \bm{0}, \forall u = 1,\cdots, U$. 
Define $\bm{v}(\omega)[z] =  \exp(-j \frac{2\pi}{\lambda}(2z\Delta_R\omega)), z = 0, \cdots,N^Z_{R} $, and 
 $\bm{s}(\varphi)[y] =  \exp\left(-j \frac{2\pi}{\lambda}(-2y\Delta_R\varphi)\right),y = 0, \cdots,N^Y_{R}$.
Then, the angles can be estimated in a manner similar to the MUSIC method by minimizing the following cost function as 
\begin{equation}\label{MUSIC}
	f(\omega,\varphi) = \bm{b}^H(\omega,\varphi) \bm{\Pi}_Q \bm{b}(\omega,\varphi), 
\end{equation}
where $\bm{b}(\omega, \varphi) =  \bm{v}(\omega) \otimes \bm{s}(\varphi)$.

Directly searching the two-dimensional angle space  requires a significant computation cost. 
In the following, we provide a low-complexity algorithm to solve the angle estimation problem.
Then, we formulate the AoA estimation problem as 
\begin{subequations}\label{PromAoA}
	\begin{align}
		\underset{\omega, \varphi}{\text{min} }  
		&\quad  f(\omega,\varphi) =  (\bm{v}(\omega)\otimes\bm{s}(\varphi))^H\bm{\Pi}_Q (\bm{v}(\omega)\otimes\bm{s}(\varphi))  \label{objAoA}  \\
		\text{s.t.} & \quad \bm{e}^H \bm{s}(\varphi) =1,\label{AoAst}
	\end{align}
\end{subequations} 
where $\bm{e} = [1, 0,\cdots,0]^T$, and the constraint (\ref{AoAst}) is introduced to avoid the trivial solution $\bm{s}(\varphi) = \bm{0}$.
Then, the objective function is reformulated as 
\begin{align}\label{fwphi}
	f(\omega,\varphi) &=  \bm{s}(\varphi)^H[ \bm{v}(\omega)^H\otimes \bm{I}_{N^Y_{R}+1}]\bm{\Pi}_Q[\bm{v}(\omega)\otimes \bm{I}_{N^Y_{R}+1}]\bm{s}(\varphi)\\
	&= \bm{s}(\varphi)^H\bm{\Theta}(\omega)\bm{s}(\varphi), 
\end{align}
where $\bm{\Theta}(\omega) =[ \bm{v}(\omega)^H\otimes \bm{I}_{N^Y_{R}+1}]\bm{\Pi}_Q[\bm{v}(\omega)\otimes \bm{I}_{N^Y_{R}+1}]$.

We construct the following Lagrange function 
\begin{equation}
	\mathcal{L} = \bm{s}(\varphi)^H\bm{\Theta}(\omega)\bm{s}(\varphi)   - \lambda (\bm{e}^H \bm{s}(\varphi) -1),
\end{equation}
where $\lambda$ is the introduced Lagrange multiplier.
Then, one obtains 
\begin{equation}
	\frac{ \partial \mathcal{L} }{\partial  \bm{s}(\varphi)} = 2\bm{\Theta}(\omega)\bm{s}(\varphi)  -\lambda \bm{e}.
\end{equation}
We can infer that $\bm{s}(\varphi) = \zeta \bm{\Theta}^{-1}(\omega)\bm{e}$, where $\zeta$ is a non-zero constant.
As $\bm{e}^H \bm{s}(\varphi) =1$, we have 
$\zeta =  (\bm{e}^H\bm{\Theta}^{-1} (\omega)\bm{e})^{-1}$.
Then, one obtains
\begin{equation} \label{varphi}
	\bm{s}(\varphi) = \frac{\bm{\Theta}^{-1}(\omega)\bm{e}}{ \bm{e}^H\bm{\Theta}^{-1} (\omega)\bm{e}}.
\end{equation}

Then, substituting (\ref{varphi}) into the objective function (\ref{fwphi}), Problem (\ref{PromAoA}) is transformed into 
\begin{equation}\label{omegaMinOpt}
	\underset{\omega}{\text{min} }  
	\quad f(\omega) =  \frac{1}{ \bm{e}^H\bm{\Theta}^{-1} (\omega)\bm{e}}.
\end{equation}
Then, the optimal solution to (\ref{omegaMinOpt}) is 
\begin{equation}\label{OmegaOpt}
	\omega^* = \arg \max\bm{e}^H\bm{\Theta}^{-1} (\omega)\bm{e}.
\end{equation}

\textit{Remark 4}: 
In practice, two-layer grids can be used for searching the optimal solution $\omega^*$.
To be specific, the grid with large step-size is used for the preliminary coarse search.
Then, based on the preliminary coarse results, the finer grid is then applied to obtain the fine results.
Furthermore, due to the presence of noise, the energy peak will leakage to adjacent grid points.
Therefore, to enhance the position accuracy, it is suggested to take several largest values as estimation candidates, and then use the clustering method, such as the K-mean method\cite{yadav2013review}, to obtain a more accurate estimation.

Recall that  $\omega_u = \sin\phi_{u}$ and $\varphi_u = \sin\theta_{u}\cos\phi_{u}$, where $\phi_{u}$ and $\theta_{u}$ represent the vertical and azimuth  AoAs of UE $u$, respectively.
Then, we can search the interval $\omega_u \in [-1, 1]$, and obtain $U$ largest peaks of the $(1,1)$-th element of $\bm{\Theta}^{-1} (\omega)$.
These $U$ peaks correspond to the  vertical  AoAs of UEs, denoted as $[\hat{\omega}_1, \hat{\omega}_2, \cdots,\hat{\omega}_U]$.
Then, for the composite angle $\varphi_u$, according to (\ref{varphi}), we can obtain the $U$ vectors $[\bm{\hat{s}}(\varphi_1), \bm{\hat{s}}(\varphi_2), \cdots,\bm{\hat{s}}(\varphi_U)]$.

Define 
\begin{equation}\label{qu}
	\bm{\hat{q}}_u = \angle\bm{\hat{s}}(\varphi_u) = [0, \frac{4\pi}{\lambda}\Delta_R\varphi_u, \cdots, \frac{4\pi}{\lambda}\Delta_R N_R^Y\varphi_u ]^T,
\end{equation}
and vector $\bm{p} = [0, \frac{4\pi}{\lambda}\Delta_R,\cdots, \frac{4\pi}{\lambda}\Delta_R N_R^Y]^T$.
Then, we utilize the LS estimator to obtain the estimation of $\varphi_u$ as
\begin{equation}\label{phiLS}
	\hat{\varphi}_u ^{LS} = (\bm{p}^T\bm{p})^{-1}\bm{p}^T\bm{\hat{q}}_u.
\end{equation}

\subsection{Estimation of Distances}

With the estimated AoAs $\{\hat{\varphi}_u\}$ and $\{\hat{\omega}_u\}$, we reformulate the array response as 
\begin{equation}
	\bm{a}_{R}(\hat{\omega}_u,\hat{\varphi}_u ,d_{R,u}^0)  =  \text{diag}\{\bm{p}_u(\hat{\omega}_u,\hat{\varphi}_u ) \}\bm{q}_u(\hat{\omega}_u,\hat{\varphi}_u ,d_{R,u}^0),
\end{equation}
where 
$\bm{p}_u(\hat{\omega}_u,\hat{\varphi}_u )= \left[\exp\left(-j \frac{2\pi}{\lambda}J_{-\frac{N_R-1}{2}}(\hat{\omega}_u,\hat{\varphi}_u) \right), \cdots, \right.$\\$\left.
\exp\left(-j \frac{2\pi}{\lambda}J_{\frac{N_R-1}{2}}(\hat{\omega}_u,\hat{\varphi}_u)\right) \right]^T$.  Also,
$$\bm{q}_u(\hat{\omega}_u,\hat{\varphi}_u ,d_{R,u}^0) = \left[\exp\left(-j \frac{2\pi}{\lambda d_{R,u}^0} q_{-\frac{N_R-1}{2}}\right), \cdots,\exp\left(-j \frac{2\pi}{\lambda d_{R,u}^0} q_{\frac{N_R-1}{2}}\right)\right]$$ with the $q_m=\left( (m_z\Delta_R)^2 +(m_y\Delta_R)^2-\left(m_z\Delta_R\hat{\omega}_u - m_y\Delta_R\hat{\varphi}_u\right)^2\right)$, and index $m$ was given by (\ref{mmymz}).

Similar to (\ref{B1B2}) and (\ref{T1T2}), we divide the matrix $\bm{A}$ and its covariance matrix $\bm{R}$ into two partitions as $\bm{A}^H = [\bm{A}^H_1, \bm{A}^H_2]$ and $\bm{R}^H = [\bm{R}^H_1, \bm{R}^H_2]$ respectively, where $\bm{A}_1,\bm{R}_1 \in \mathbb{C}^{U \times N_R}$ collects the first $U$ rows and $\bm{A}_2, \bm{R}_2\in \mathbb{C}^{(N_R-U) \times N_R}$ contains the remaining $(N_R-U)$ rows.
As $\bm{A}_1$ and $\bm{R}^H_1$ are of full rank $U$, the rows of $\bm{R}_2(\bm{A}_2)$ can be expressed as a linear combination of linear independent rows of $\bm{R}_1(\bm{A}_1)$ with the  $U \times (N_R-U)$ linear operator $\bm{P} _2$.
Similar to (\ref{P1B1})-(\ref{PT1T2}), one obtains
\begin{equation} \label{PA1A2}
	\bm{P}_2^H \bm{A}_1 = \bm{A}_2,\bm{P}_2^H \bm{R}_1 = \bm{R}_2.
\end{equation}
Then, $\bm{P} _2$ is given by 
\begin{equation}
	\bm{P}_2  = (\bm{R}_1\bm{R}_1^H)^{-1}\bm{R}_1\bm{R}_2^H.
\end{equation}
Defining $\bm{Q}^H_p = [\bm{P}_2^H, -\bm{I}_{N_R- U}]$, the orthogonal projector of $\bm{A}$ is given by
\begin{align}
	\bm{Q}^H_p  \bm{A} = \bm{0},\bm{\Pi}_{p} \triangleq  \bm{Q}_p (\bm{Q}_p^H \bm{Q}_p)^{-1}\bm{Q}_p^H .
\end{align}

Then, similar to (\ref{MUSIC}), we have the following distance estimation problem 
\begin{align} 
{d^0_{R,u}}^* \!\!\!&=\underset{d_{R,u}^0}{\text{min} }  f(d_{R,u}^0) \label{dRu} \\
& = \underset{d_{R,u}^0}{\text{min} }  ( \text{diag}\{\bm{p}_u \}\bm{q}_u(d_{R,u}^0)  )^H\bm{\Pi}_{p}( \text{diag}\{\bm{p}_u \}\bm{q}_u(d_{R,u}^0) ) .  \nonumber
\end{align}
Then, the distance can be estimated by conducting one-dimensional search in the interval $[d_{min}, d_{max}]$.
The maximal $U$ largest peaks of the searching results are readily obtained as the estimated distances.

\subsection{Estimation of Locations and Channel Gains }

With the AoAs and distances, the position of UE $u$ denoted as $(x_u,y_u,z_u)$ can be obtained according to its geometric relationship with RIS position $(x_R,y_R,z_R)$, given by
\begin{align}
	& x_u = \sqrt{(1- \hat{\omega}_u^2)\hat{d}^2_{R,u} - \varphi^2_u  \hat{d}^2_{R,u}}.   \label{xu}\\
	& y_u = \varphi_u  \hat{d}_{R,u}  \label{yu}\\
	&z_u = z_R - \hat{\omega}_u\hat{d}_{R,u}   \label{zu} 
\end{align}

It is worth pointing out that , in the far-field model,  with one RIS panel, only the AoAs  can be obtained, but the 3D position of the UE cannot be obtained due to the lack of dimensions.
To obtain the channel fading coefficients $\bm{g}= [g_{1}, \cdots, g_{U}]$, we utilize the received signal $\bm{y}_t$ in the $t$-th pilot data duration, which collects the received signal with these $L$ different phase shift vectors as
\begin{equation} 
	\bm{y}_t =  \bm{G}_{RIS}\bm{A}\bm{x}_t+ \bm{n}_t \\
	= \sum_{u=1}^{U} \bm{G}_{RIS}\bm{a}_{R}({\omega}_u,{\varphi}_u,{d}_{R,u})\bm{x}_t(u)+ \bm{n}_t,
\end{equation}
where the received vector $\bm{y}_t $ has the size of $N_AL\times 1$.    

Without loss of generality, we sort the UEs according to the ascending order of their distances, i.e., $d_1\leq \cdots  \leq d_U$.
The reason is that a far UE normally experiences a severe channel fading gain.
Then, the received residual signal is initialized as $\bm{y}^{Res}_{0} = \bm{y}_t$, and the residual signal for UE $u$ is calculated as 
\begin{equation}
	\bm{y}^{Res}_{u}  =   \bm{y}_t  -  \sum_{k=1}^{u-1} \hat{g}_{k,t}\bm{G}_{RIS}\bm{a}_{R}(\hat{\omega}_k,\hat{\varphi}_k,\hat{d}_{R,k})\bm{x}_t(k). 
\end{equation}   
According to the OMP algorithm, the complex channel gain can be formulated as the solution to 
\begin{equation}
	\arg \underset{g_{u,t}}{\text{min} }  ||\bm{y}^{Res}_{u} - \hat{g}_{u,t}\bm{G}_{RIS}\bm{a}_{R}(\hat{\omega}_u,\hat{\varphi}_u,\hat{d}_{R,u})\bm{x}_t(u)||^2. 
\end{equation}
Then, the estimated channel fading parameter is
\begin{equation}\label{gut}
	\hat{g}_{u,t} = \frac{ \bm{a}_{R,u}^H \bm{y}^{Res}_{u}}{\sqrt{P_{u}^t} \bm{a}_{R,u}^H\bm{a}_{R,u}},
\end{equation} 
where $\bm{a}_{R,u} = \bm{G}_{RIS}\bm{a}_{R}(\hat{\omega}_u,\hat{\varphi}_u,\hat{d}_{R,u})$.
After we have collected $\tau$ different slot pilot data, the estimated channel fading gain for each time slot should be averaged as 
\begin{equation}
	\hat{g}_{u}   =  \frac{1}{\tau} \sum_{t=1}^{\tau} \hat{g}_{u,t}.
\end{equation}
Overall, the complex channel gain of the RIS-UE $u$ link can be obtained as 
\begin{equation}
	\hat{\bm{h}}_u = \hat{g}_{u}\bm{G}_{A,R}\bm{E}_{RIS}\bm{G}_{RIS}\bm{a}_{R}(\hat{\omega}_u,\hat{\varphi}_u,\hat{d}_{R,u}).
\end{equation}

\subsection{Complexity Analysis}

\begin{algorithm}
	\caption{Near Field Joint Channel Estimation and Localization (NF-JCEL) Algorithm}
	\begin{algorithmic}[1]\label{alg1}
		\REQUIRE   \ \ The received signal $\bm{y}_t \in N_AL \times 1$ ;  \\
		\quad  \quad \ \ pilot data $x_t$, $t = 1,\cdots,\tau$.
		
		\ENSURE    The AoAs of the RIS-UE links $\omega_u$, $\varphi_u$;  \\
		\quad  \quad \ \ The distance $d_{R,u}^0$ of the $u$-th RIS-UE link; \\
		\quad  \quad \ \   The channel fading coefficients $g_u$; \\
		\quad \quad \ \ The UE locations$(x_u,y_u,z_u)$; \\
		\quad \quad \ \  Complex channel gains $\hat{\bm{h}}_u$. \\
		
		\STATE   Estimate array covariance matrix $\hat{\bm{R}}$ according to (\ref{Rhat}) ;
		\STATE   Construct matrix $\bm{T}$ according to (\ref{ToepMat});
		\STATE   Calculate AoAs $\omega_u$ and $\varphi_u$ by solving Problem (\ref{PromAoA}), where the optimal $\omega_u^*$ is obtained according to (\ref{OmegaOpt}) and the estimated $\varphi_u$ is obtained according to (\ref{phiLS});
		\STATE   Search optimal distance ${d_{R,u}^0}^*$ by solving Problem (\ref{dRu});
		\STATE  Calculate locations$(x_u,y_u,z_u)$ according to (\ref{zu}) -(\ref{xu}); 
		\STATE  Obtain the channel fading coefficients $\bm{g}= [g_{1}, \cdots, g_{U}]$ by using the OMP method;
	\end{algorithmic}
\end{algorithm}

In summary, a NF-JCEL algorithm is proposed to estimate the UE locations and the complex channel coefficients. 
The detailed algorithm is presented in Algorithm \ref{alg1}.
The proposed NF-JCEL algorithm solves three sub-problems sequentially: 
1) the estimation of array covariance matrix based on the LS estimation;
2) the estimation of locations, including the AoAs and distance;
3) the channel fading coefficients estimation.
Therefore, the complexity of the NF-JCEL algorithm can be evaluated from these three parts, respectively. 

The complexity of step 1 mainly lies in the matrix calculation. The matrix $\bm{G}_{RIS}$ has size $N_AL \times N_R$, and the complexity of product $\bm{G}_{RIS} ^H\bm{G}_{RIS}$ is $\mathcal{O}(N_R^2N_AL)$. Its inversion requires the complexity of $\mathcal{O}(N_R^3)$. Thus, the calculation of $\bm{H}^{\dag}_{RIS}$ has the total complexity of $\mathcal{O}(N_R^3 + 2N_R^2N_AL)$. Then, the complexity of LS estimation $\widehat{\bm{A}\bm{x}}_t$ is given by $\mathcal{O}(N_R^3 + 2N_R^2N_AL + N_RN_AL)$.
Finally, the complexity of array covariance matrix estimation is $\mathcal{O}(\tau (N_R^3 + 2N_R^2N_AL + N_RN_AL +N_R)) = \mathcal{O}(\tau (N_R^3 + 2N_R^2N_AL))$.

The complexity of UE localization mainly lies in the orthogonal projectors and optimization objectives. 
Similarly, according to (\ref{POp}), $\bm{P}_1$ is calculated with complexity $\mathcal{O}( U^3+ 2U^2N_b + (N_b-U)UN_b)$, and the complexity of calculating $\bm{\Pi}_Q$ is $\mathcal{O}((N_b-U)^3 + 2(N_b-U)^2N_b)$.
Then, the complexity of calculating $\bm{\Theta}^{-1} (\omega)$ is $\mathcal{O}( (N_R^Y+1)N_b^2 + (N_R^Y+1)^2N_b + (N_R^Y+1)^3 + (N_R^Y+1)^2 +N_R^Y) = \mathcal{O}( (N_R^Y+1)N_b^2 + (N_R^Y+1)^2N_b)$.
Consequently, the total complexity of estimating $\omega_u$ is $\mathcal{O}(N_{\omega}((N_R^Y+1)^2N_b^2 + (N_R^Y+1)^3N_b))$, where $N_{\omega}$ denotes the number of searching grids for solving Problem (\ref{OmegaOpt}).
In addition, the complexity of estimating ${\phi}_u$ is $\mathcal{O}(2(N_Y+1))$.
Furthermore, the complexity of solving Problem (\ref{dRu}) is given by $\mathcal{O}(N_d(2N_R^2 + N_R))$, where $N_{d}$ denotes the number of searching grids.
Overall, the complexity of UE localization is $\mathcal{O}(UN_{\omega}((N_R^Y+1)^2N_b^2 + (N_R^Y+1)^3N_b)+2(N_Y+1)+N_d(2N_R^2 + N_R))$.
Finally, the complexity of channel fading estimation is $\mathcal{O}(2\tau N_AU)$.
Then, the overall complexity of NF-JCEL algorithm is given by $\mathcal{O}(\tau (N_R^3 + 2N_R^2N_AL) + UN_{\omega}((N_R^Y+1)^2N_b^2 + (N_R^Y+1)^3N_b)+2(N_Y+1)+N_d(2N_R^2 + N_R) + 2\tau N_AU)$.

\section{Simulation results}\label{sumSec}

In this section, some representative simulation results are presented for performance validation.
In the considered scenario, UEs are distributed in a $5$ m $\times$ $5$ m square area.
The AP is installed on the wall at $(1.3,0,2.7)$ with $25$ antenna elements, and the separation between antenna elements is set as $2$ mm.
The RIS is installed on the wall at $(0,1,2.5)$ with $3.33$ mm separations of reflecting elements.
The noise power is set to $-120$ dBm.
The transmission carrier frequency is $90$ GHz, $G_t= 40$ dBi, $G_r= 50$ dBi and transmit power is $27$ dBm.
Unless specified otherwise, the RIS panel is equipped with $11\times101 = 1111$ reflecting elements with Fraunhofer distance of $17.58$ m, the number of RIS training phase shift vectors is $S = 150$, the number of pilot data is $\tau = 20$.
All results are obtained by averaging over $200$ random generations of UE locations.
For performance comparison, we utilize the conventional far-field model with the MUSIC method as the performance benchmark, which is labeled as ``FF''.
Note that the UEs are supposed to be on the XOY plane, i.e., $z_u =0$, as in this case, the far-field method can only obtain the 2D localization.
The proposed NF-JCEL algorithm is labeled as ``Prop''.

	\begin{figure}
		\centering
		\includegraphics[width=0.4\textwidth]{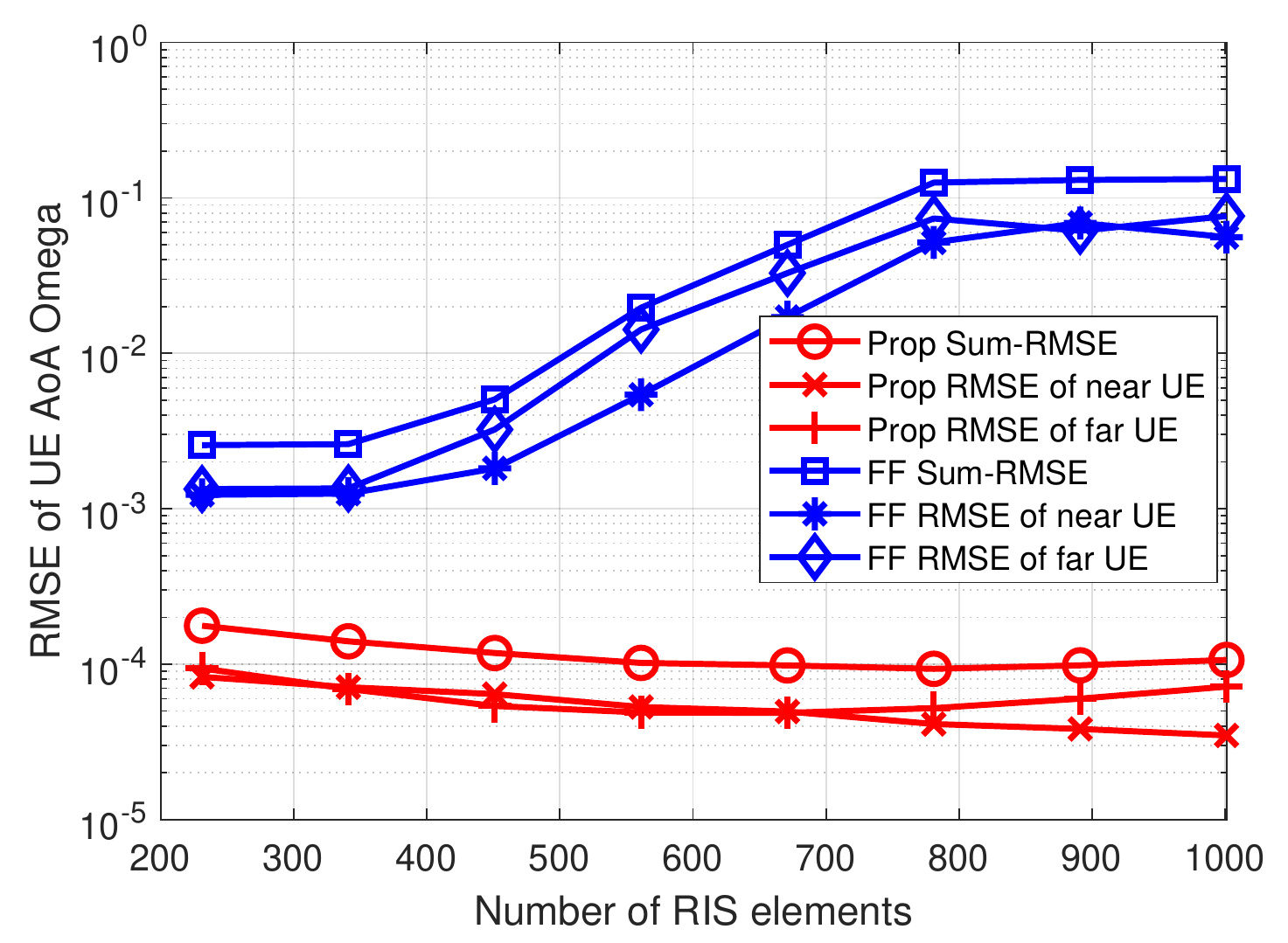}
		\caption{RMSE of AoA $\omega$ versus number of RIS elements.}
		\label{Fig_Omega_RISNum}
		\vspace{-1em}
	\end{figure}

	\begin{figure}
		\centering
		\includegraphics[width=0.4\textwidth]{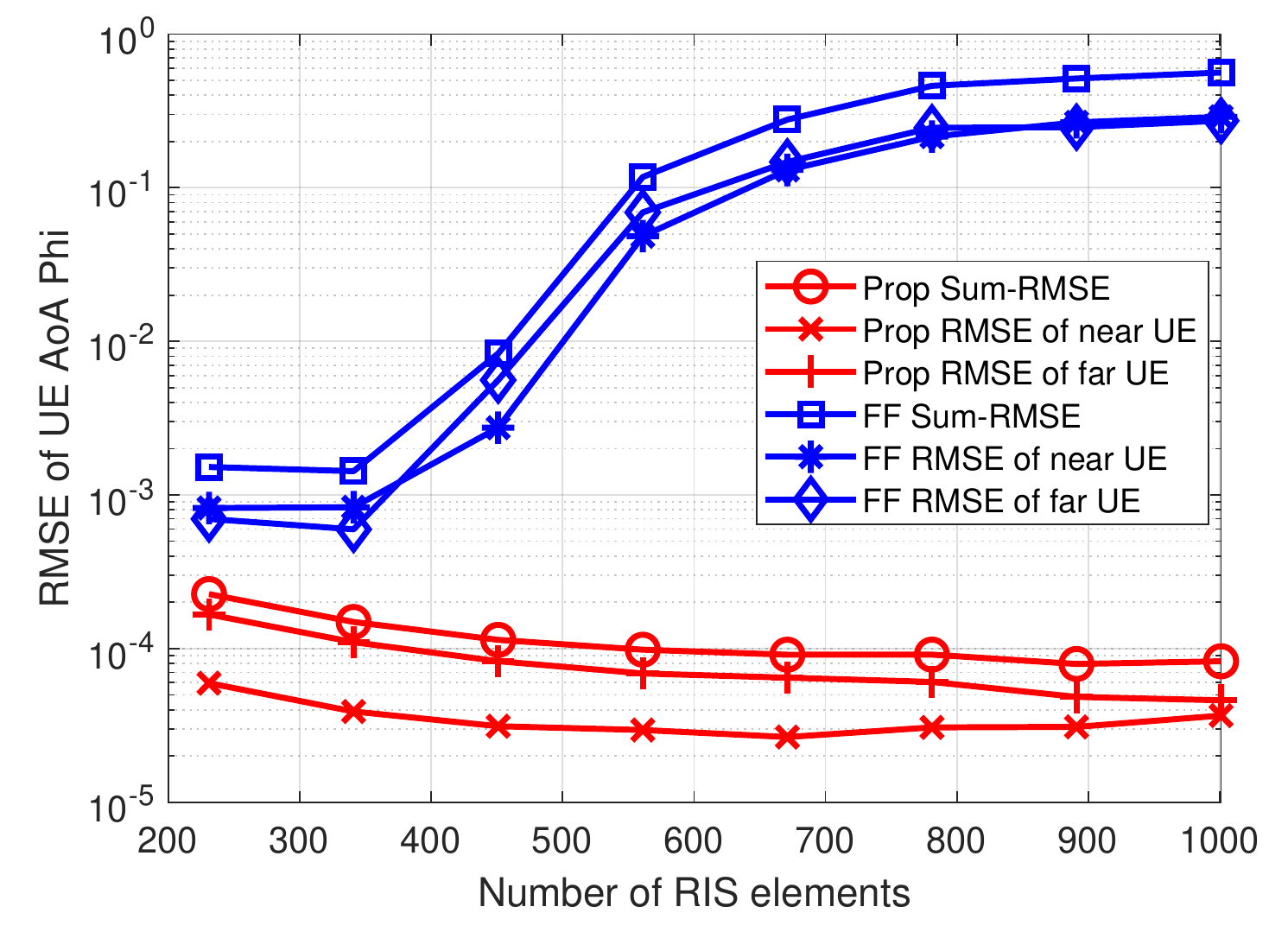}
		\caption{RMSE of AoA $\varphi$ versus number of RIS elements.}
		\label{Fig_Phi_RISNum}
		\vspace{-1em}
	\end{figure}

	\begin{figure}
		\centering
		\includegraphics[width=0.4\textwidth]{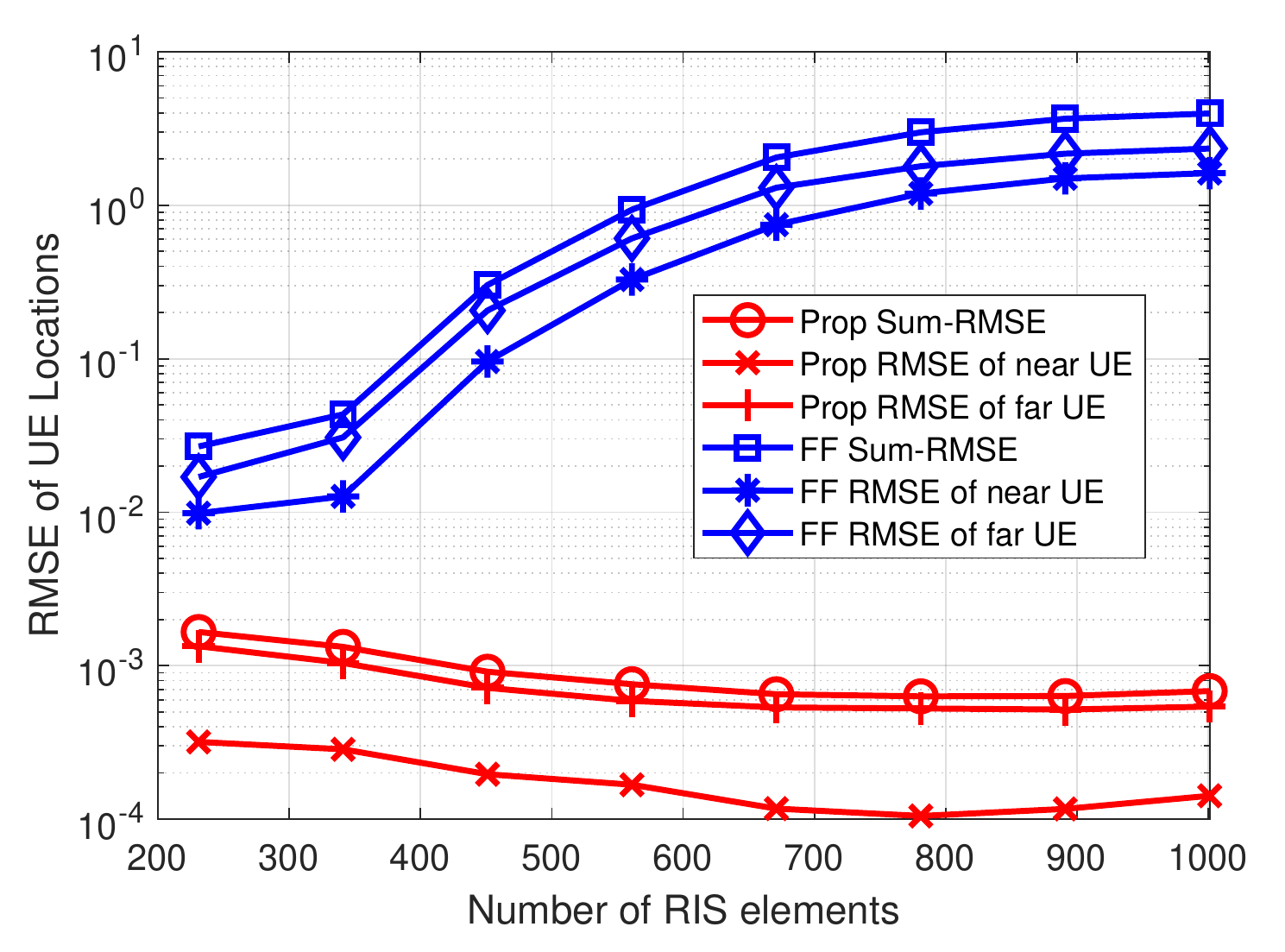}
		\caption{RMSE of location versus number of RIS elements.}
		\label{Fig_Loc_RISNum}
		\vspace{-1em}
	\end{figure}

	\begin{figure}
		\centering
		\includegraphics[width=0.4\textwidth]{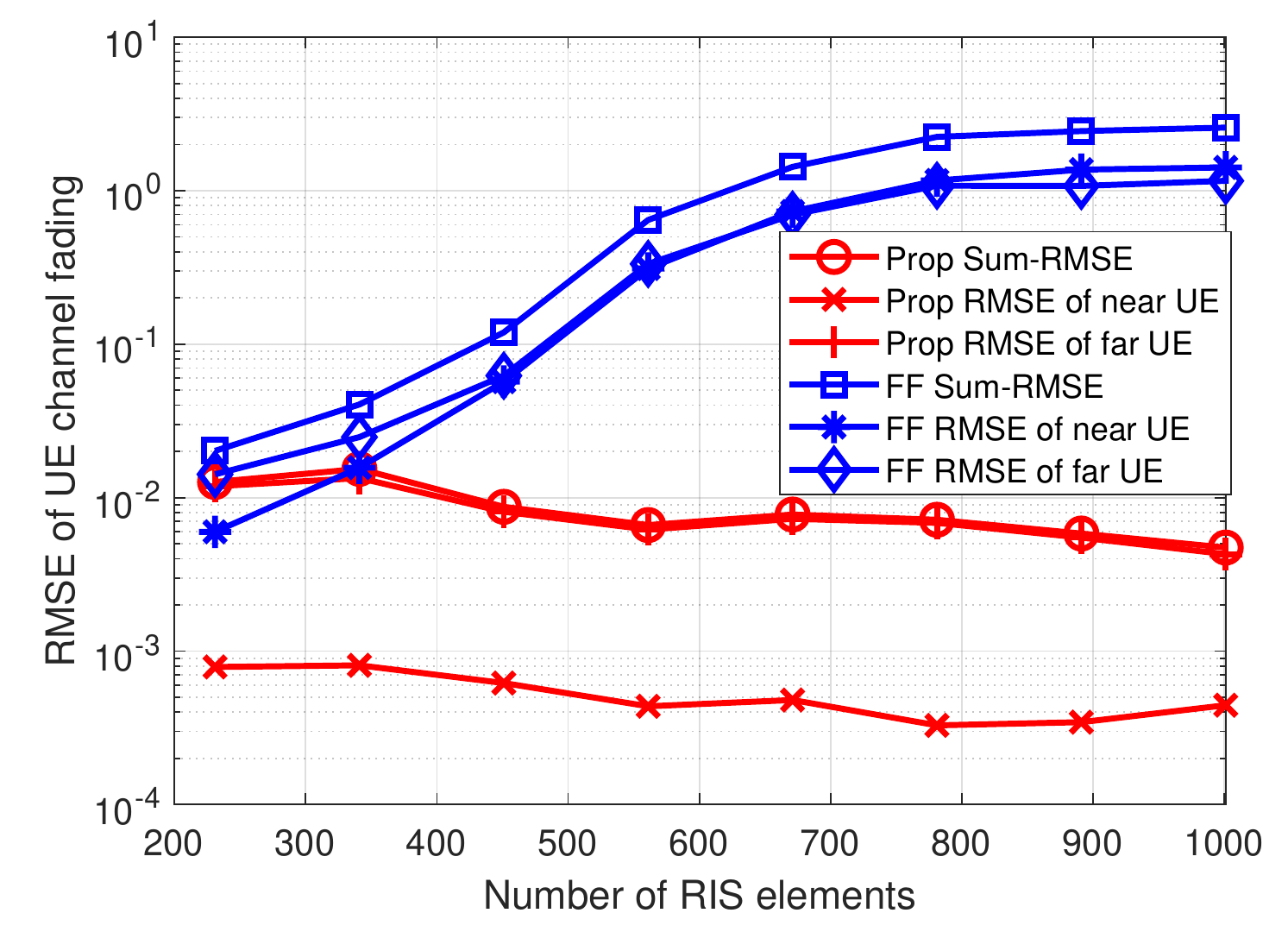}
		\caption{RMSE of channel gain versus number of RIS elements.}
		\label{Fig_H_RISNum}
		\vspace{-1em}
	\end{figure}

\subsection{Impact of the Number of RIS Elements}

Fig. \ref{Fig_Omega_RISNum}-Fig.\ref{Fig_H_RISNum} show the RMSE performances as functions of the numbers of reflecting elements, where the Fraunhofer distance corresponding to the number of reflecting elements is from $0.4823$ m to $8.1$ m.
The number of UE is $2$, one is a close UE and the other is a far UE, resulting in three RMSE curves: the respective RMSEs of these two UEs and the sum of two RMSEs, which are labeled as ``RMSE of near UE'', ``RMSE of far UE'' and ``Sum-RMSE'', respectively.
Unless specified otherwise, the same configuration and labels are adopted in the following subsections.

Fig. \ref{Fig_Omega_RISNum} shows the RMSEs of the estimated AoA $\omega_u = \sin\phi_{u}$.
It can be seen from Fig. \ref{Fig_Omega_RISNum}  that the sum-RMSE of the proposed algorithm decreases with increasing the number of reflecting elements, while the sum-RMSE of the far-field algorithm increases with the number of reflecting elements.
This means that the near-field effect increases with increasing the number of reflecting elements, i.e. the panel size of RIS.
To be more specific, the channel model approximation error of the far-field algorithm increases with the number of reflecting elements.
Fig. \ref{Fig_Phi_RISNum} shows the RMSEs of the estimated AoA $\varphi_u = \sin\theta_{u}\cos\phi_{u}$.
From Fig. \ref{Fig_Phi_RISNum}, similar trends in the sum RMSE can be observed to those from Fig. \ref{Fig_Omega_RISNum}.

Fig. \ref{Fig_Loc_RISNum} shows the localization performance in terms of RMSE. 
It can be seen from Fig. \ref{Fig_Loc_RISNum} that the proposed algorithm outperforms the corresponding far-field cases, and the performance gap increases with the number of reflecting elements.
This implies that the proposed algorithm can fully leverage the information provided in the near-field scenario. 
The RMSE of the near UE is less than that of the far UE for both far-field and proposed near-field algorithms, as the localization RMSE is also heavily affected by distance-dependent fading.
Fig. \ref{Fig_H_RISNum} shows the RMSE performance of the channel fading coefficients.
Similar trends to Fig. \ref{Fig_Loc_RISNum} can be seen from Fig. \ref{Fig_H_RISNum}.
In Fig. \ref{Fig_H_RISNum}, it is observed that the RMSE of the proposed algorithm is dominated by the far UEs, as the channel fading is more severe for the longer transmission distance.

\subsection{Impact of Number of UEs}

	\begin{figure}
		\centering
		\includegraphics[width=0.4\textwidth]{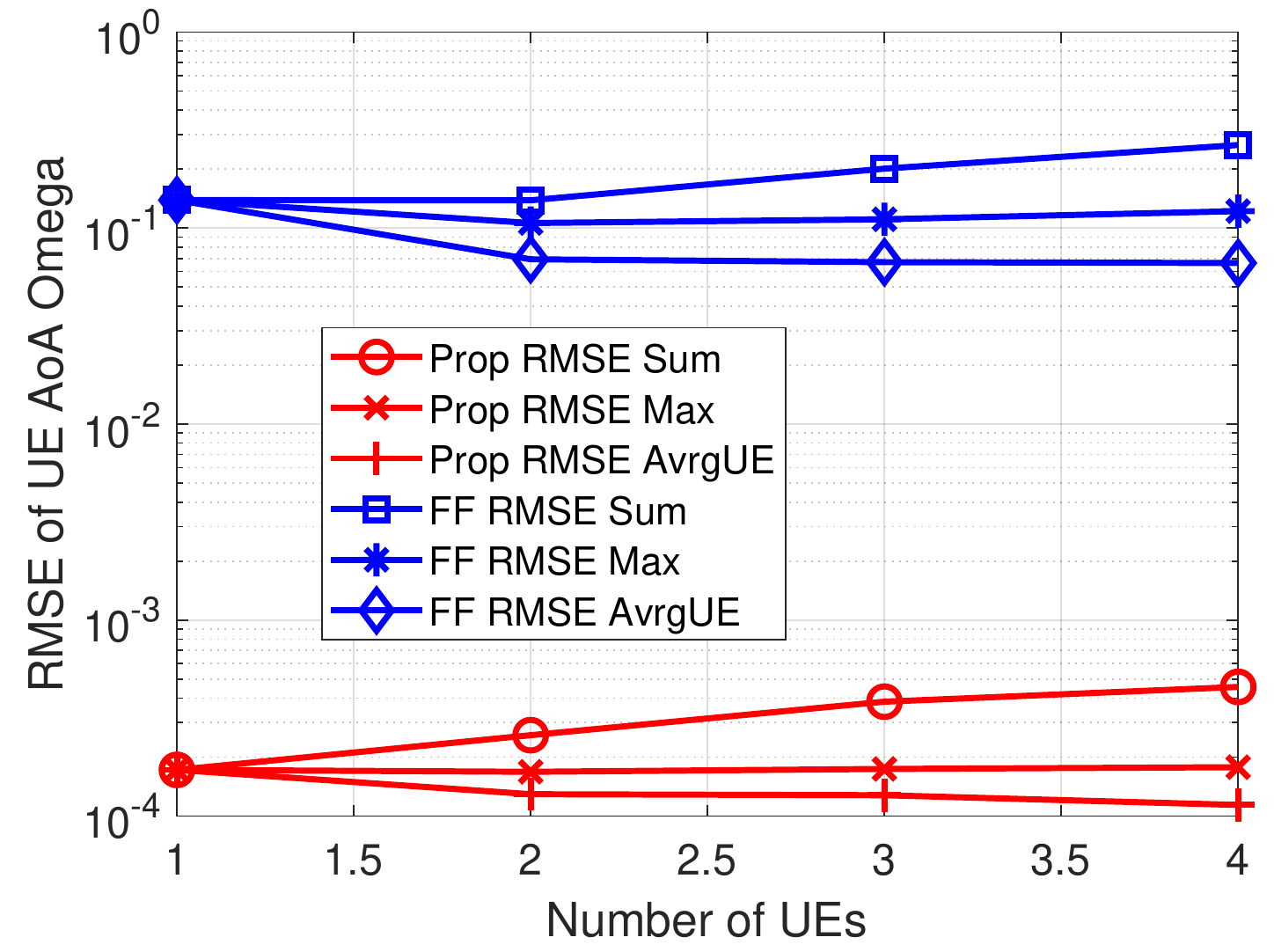}
		\caption{RMSE of AoA $\omega$ versus number of UEs.}
		\label{Fig_Omega_UENum}
		\vspace{-1em}
	\end{figure}

	\begin{figure}
		\centering
		\includegraphics[width=0.4\textwidth]{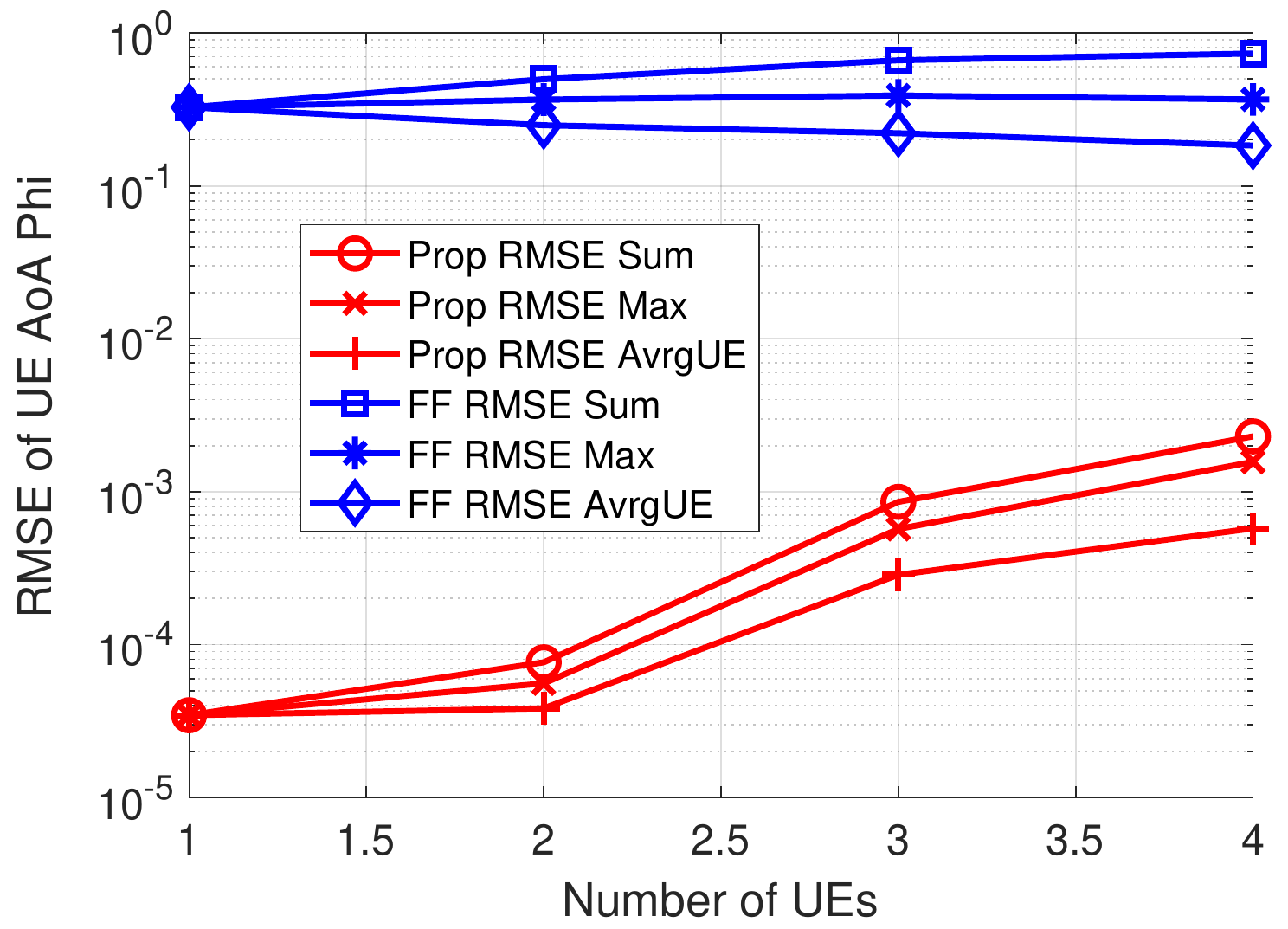}
		\caption{RMSE of AoA $\varphi$ versus number of UEs.}
		\label{Fig_Phi_UENum}
		\vspace{-1em}
	\end{figure}

Fig. \ref{Fig_Omega_UENum} shows the RMSE of $\omega_u = \sin\phi_{u}$ versus the number of UEs, where the number of UEs gradually increases by adding UEs moving away from the RIS panel.
The sum of RMSEs for all UEs, the worst RMSE among all UEs, and the average RMSE per UE are shown, which are labeled as  ``RMSE Sum'', ``RMSE Max'' and ``RMSE AvrgUE'', respectively.
It can be seen that the proposed algorithm always performs better than the far-field case.
The sum RMSEs of the proposed algorithm increase with the number of UEs, as more UEs need to be located.
In addition, the sum-RMSE of AoA estimation is dominated by the worst RMSE in the proposed algorithm, and the performance gap between the UE's average RMSE and the sum-RMSE increases with the number of far-away UEs.
Meanwhile, the``RMSE Max'' and ``RMSE AvrgUE'' of the far-field algorithm slightly decreases with increasing number of UEs.
The reason is that more far-away UEs are added in the simulation, as the increased number of distant UEs do not result in more severe near-field effects.

Fig. \ref{Fig_Phi_UENum} shows the RMSE of $\varphi_u = \sin\theta_{u}\cos\phi_{u}$.
Similar trends of the far-field algorithm can be seen from Fig. \ref{Fig_Phi_UENum} to those from Fig. \ref{Fig_Omega_UENum}. 
Also, the sum RMSEs of the proposed algorithm increase with the number of UEs, and the main reason is that UEs' separations in the angular domain decreases with the number of UEs.

	\begin{figure}
	\centering
	\includegraphics[width=0.4\textwidth]{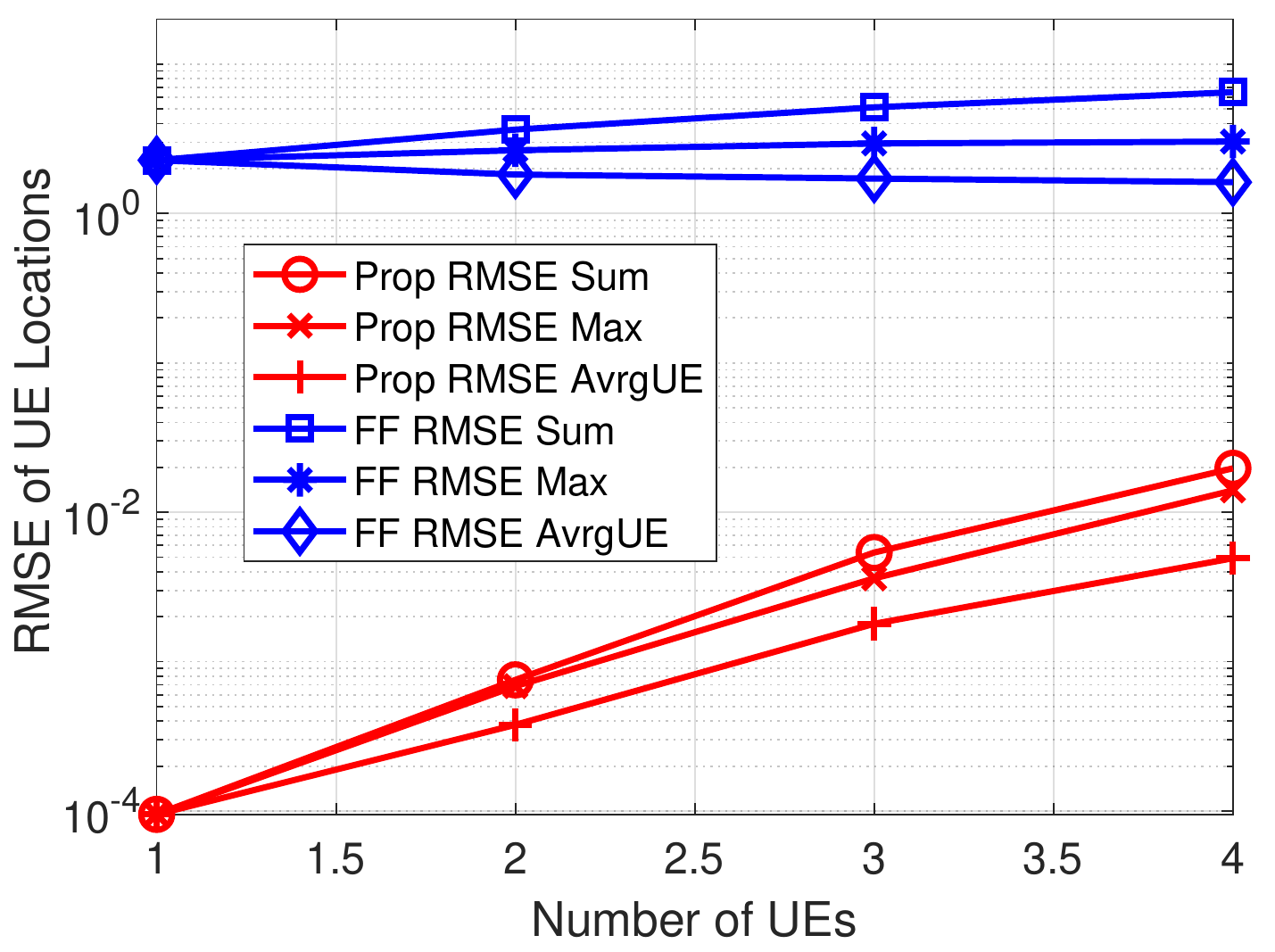}
	\caption{RMSE of location versus number of UEs.}
	\label{Fig_Loc_UENum}
	\vspace{-1em}
\end{figure}

Fig. \ref{Fig_Loc_UENum} shows the localization performance.
The localization RMSEs obtained by the proposed algorithm increase with the number of UEs, but the proposed algorithm shows better performance than the far-field algorithm.
Meanwhile, the``RMSE Max'' and ``RMSE AvrgUE'' of the far-field algorithm decreases slightly.
This is consistent with Fig. \ref{Fig_Omega_UENum} and Fig. \ref{Fig_Phi_UENum}.
Furthermore, the performance gap between the far-field algorithm and the proposed algorithm decreases.
This is because the added UE is gradually far away from RIS panels so that the near field effect gradually becomes insignificant.

\subsection{Impact of Transmit Power}
	
	\begin{figure}
		\centering
		\includegraphics[width=0.4\textwidth]{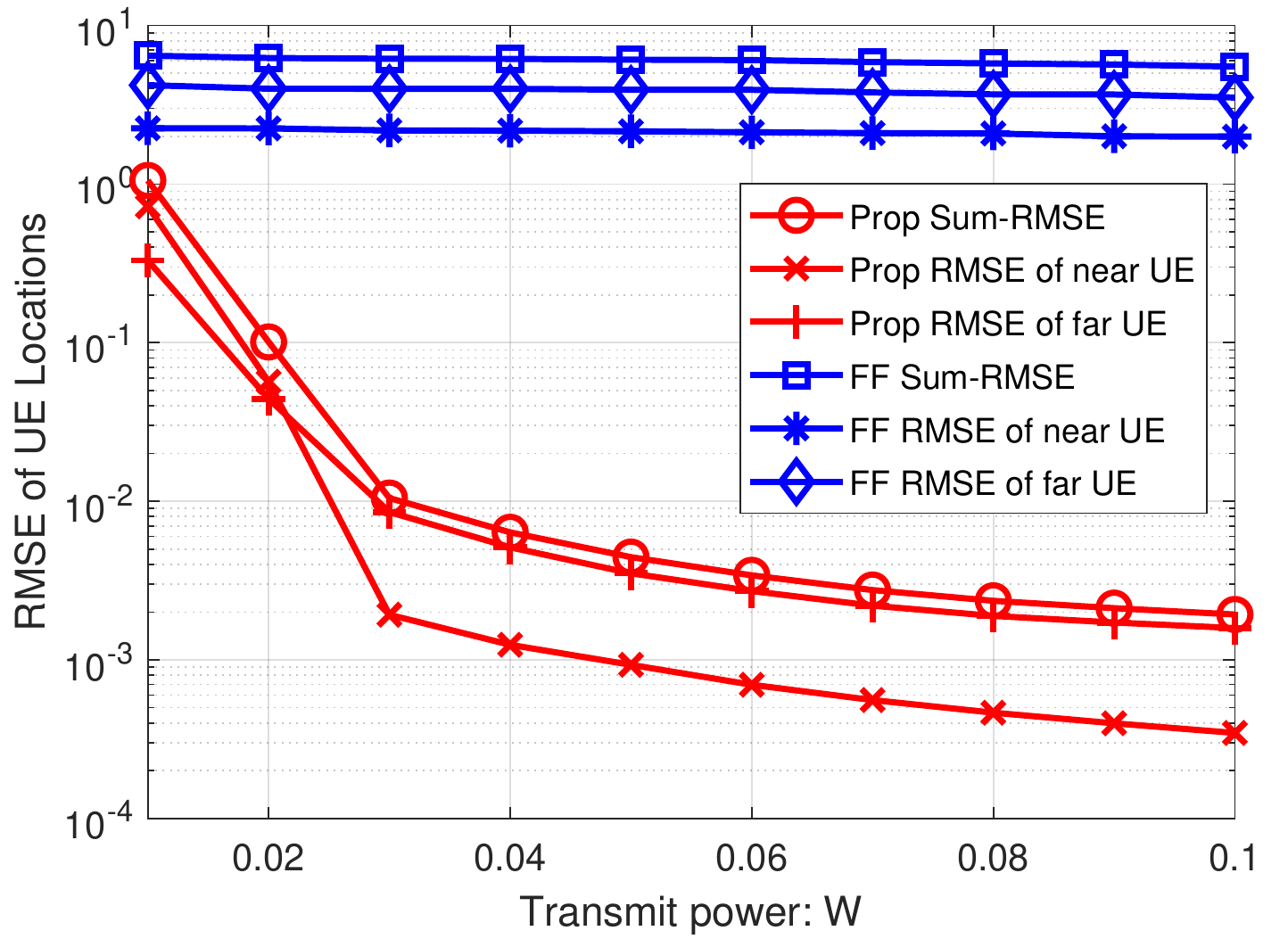}
		\caption{RMSE of location}
		\label{Fig_loc_Pw}
		\vspace{-1em}
	\end{figure}

	\begin{figure}
		\centering
		\includegraphics[width=0.4\textwidth]{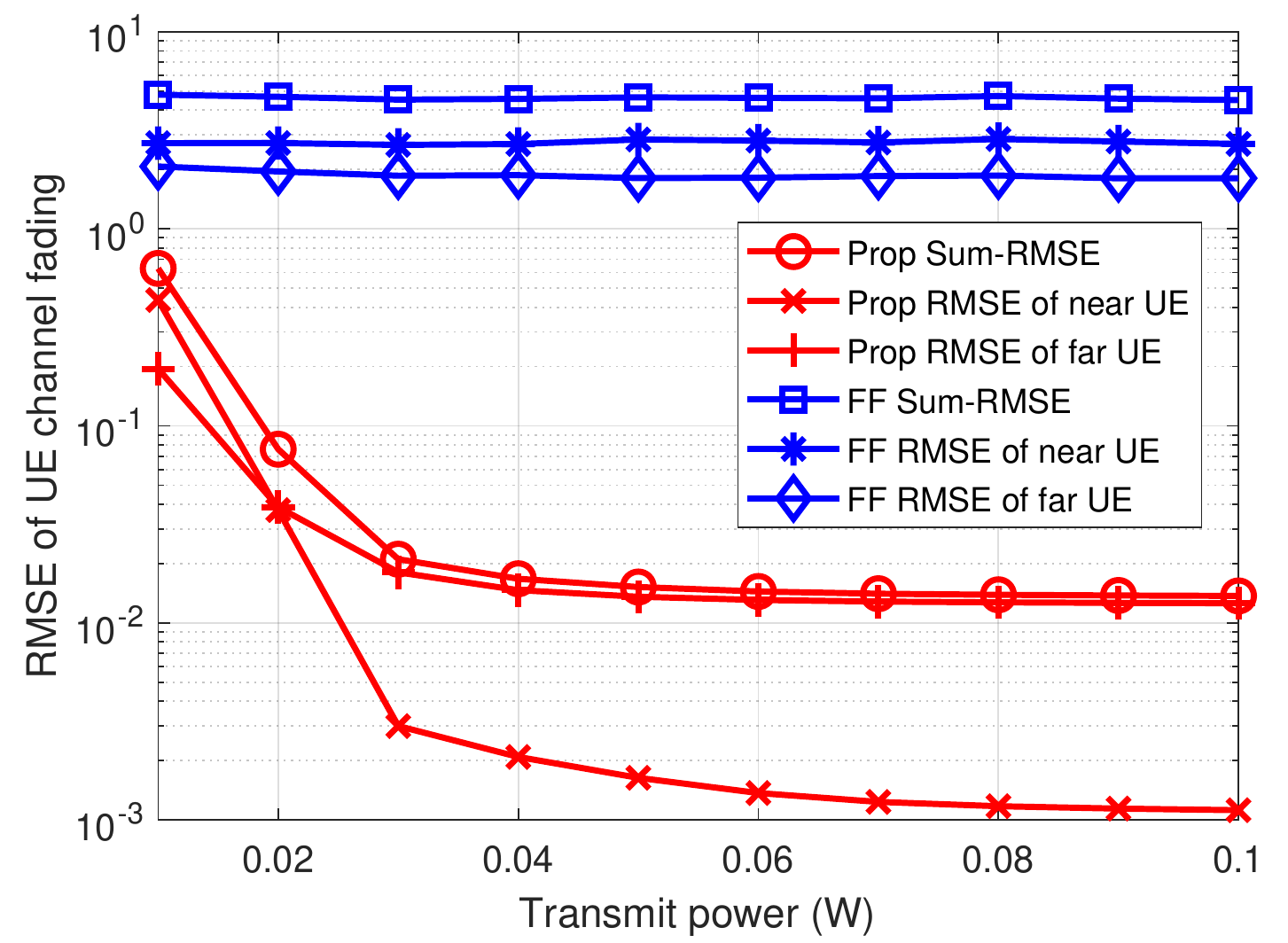}
		\caption{RMSE of channel fading cofficients}
		\label{Fig_H_Pw}
		\vspace{-1em}
	\end{figure}

Fig.\ref{Fig_loc_Pw}and Fig. \ref{Fig_H_Pw} show the impact of the transmit power on the RMSE performance.
It can be seen from Fig. \ref{Fig_loc_Pw} that the proposed algorithm outperforms the far-field algorithm for all cases.
The RMSEs of the proposed algorithm decrease with the transmit power, while the RMSE of the far-field algorithm keeps flat.
The sum RMSE of the proposed algorithm is dominated by that of the far UE, and the RMSE of the near UE decreases faster than the far UE.
Fig. \ref{Fig_H_Pw} shows the RMSE of the channel fading coefficients, and similar trends to those in Fig. \ref{Fig_loc_Pw} are illustrated.

\subsection{Impact of Pilot Overhead}

\begin{figure}
		\centering
		\includegraphics[width=0.4\textwidth]{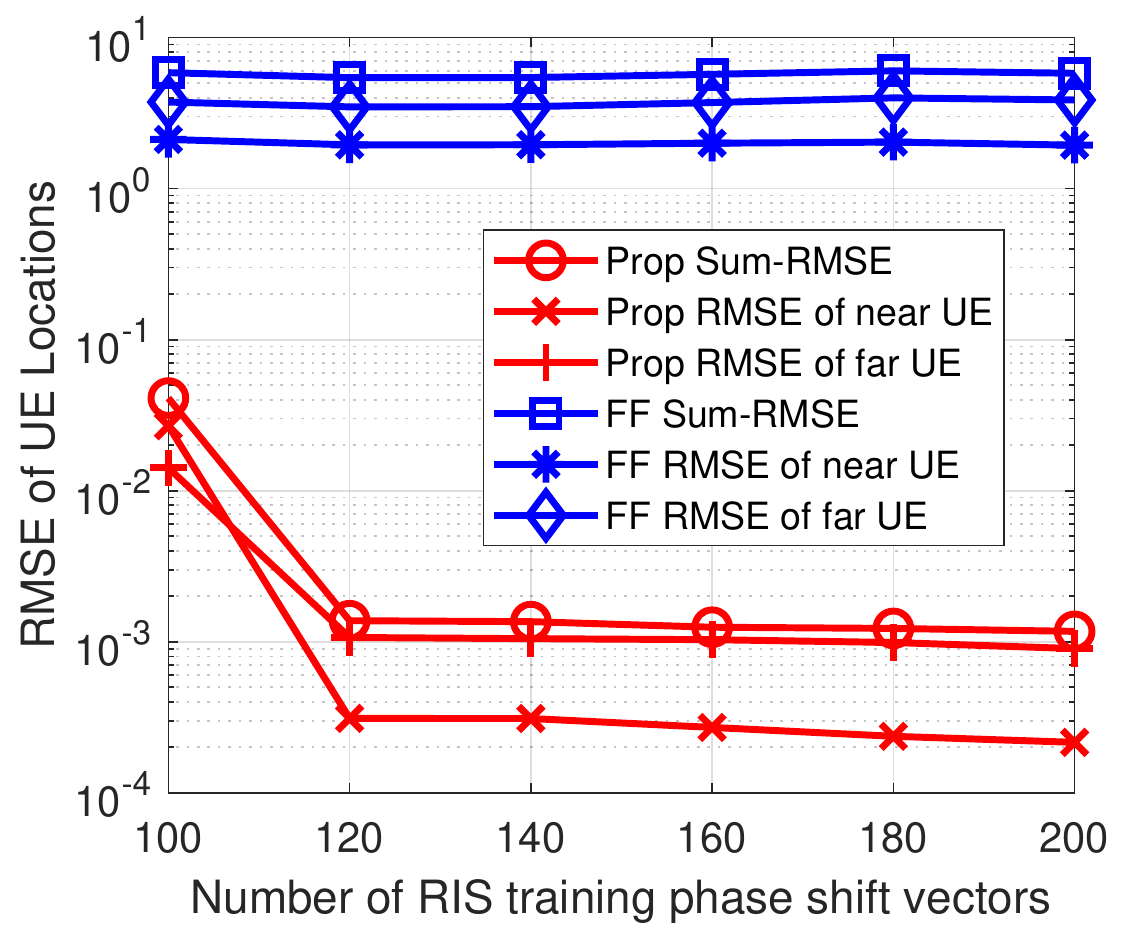}
		\caption{RMSE of location versus the number of RIS training phase shift vectors.}
		\label{Fig_loc_RISSlot}
		\vspace{-1em}
	\end{figure}

\begin{figure}
		\centering
		\includegraphics[width=0.4\textwidth]{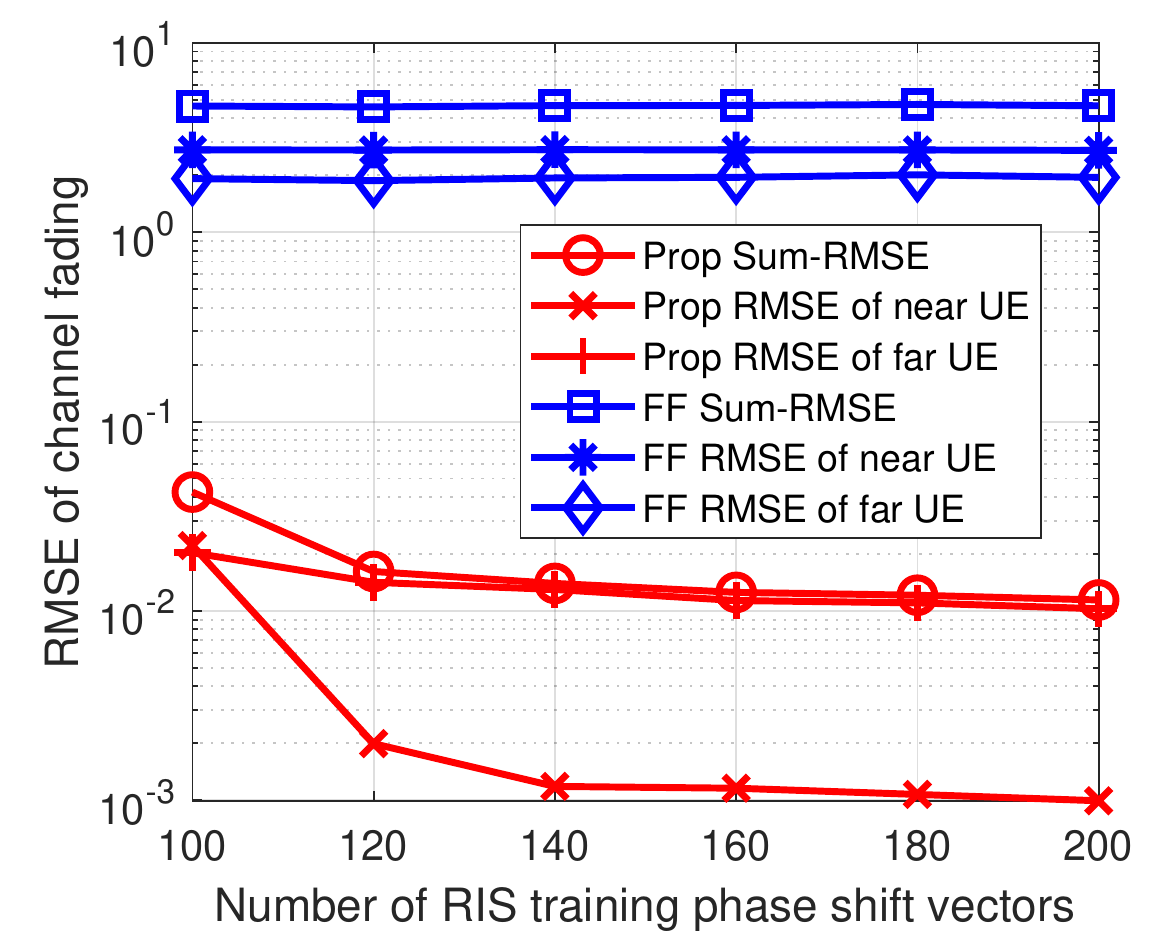}
		\caption{RMSE of channel fading coefficients versus the number of RIS training phase shift vectors.}
		\label{Fig_H_RISSlot}
		\vspace{-1em}
	\end{figure}

Fig. \ref{Fig_loc_RISSlot} and Fig. \ref{Fig_H_RISSlot} show the impact of the number of RIS training phase shift vectors, i.e. $S$, on the RMSE performance.
It can be seen from Fig. \ref{Fig_loc_RISSlot} that the proposed algorithm outperforms the far-field algorithm.
The RMSE of locations obtained by the proposed algorithm decreases rapidly when the number of training phase shift vectors increases to $120$.
Then, the decent speeds of RMSE slow down for the proposed algorithm.
Meanwhile, the RMSE of the far-field algorithm keeps flat with the number of training phase shift vectors.
Fig. \ref{Fig_H_RISSlot} shows the RMSE of the channel fading coefficients.   
This is consistent with  Fig. \ref{Fig_loc_RISSlot}.          
The reason can be explained as follows.
With more training phase shift vectors,  i.e.,  the increased number of received pilot signals, the proposed algorithm can fully exploit the near-field information, while the performance of the far-field algorithm cannot be improved.

\section{Conclusion}

In this paper, we have investigated the localization and CSI estimation scheme for the near-field subTHz system with RIS.
The following conclusions are drawn: 
\begin{itemize}
	\item The proposed NF-JCEL algorithm shows attractive performance in terms of localization and CSI estimation RMSE, especially for the near-field transmission introduced by the XL-RIS panel. Meanwhile, the conventional far-field algorithm shows severe performance degradation in the near-field channel.
	
	\item The complexity in the near-field CSI estimation is highly dependent on the array steering vector formulation, where the UE distance and the AoAs are different by reflecting elements and jointly coupled together.
	However, this also leads to higher resolution accuracy.
	Thus, the near-field channel model can provide 3D localization with a single RIS panel, which is not applicable for the far-field model with one RIS panel in the same case.
	
	\item  The localization accuracy is also highly dependent on the angle separations between different UEs, which can be improved with a large RIS panel with more elements.
	However, without considering the near-field effect, the more reflecting elements cannot bring more benefits but severe performance degradation.
	Thus, it is a necessity to consider the spherical wavefront feature for the high precision localization system.
\end{itemize}

\bibliographystyle{ieeetran}
\bibliography{Reference}

\end{document}